# Landau, Ginzburg, Devonshire and others

Arkady P. Levanyuk[a,*], I. Burc Misirlioglu[b,c], and M. Baris Okatan[d]

[a]*University of Washington, Department of Physics, Seattle, WA USA;* [b]*Faculty of Engineering and Natural Sciences, Sabanci University, Orhanli/Tuzla, Istanbul, Turkey;* [c]*Sabanci University Nanotechnology Research and Application Center, Orhanli/Tuzla, Istanbul, Turkey;* [d]*Department of Materials Science and Engineering, Izmir Institute of Technology, Gulbahce/Urla, Izmir, Turkey*

[*] Author to whom correspondence should be addressed: A. P. Levanyuk, levanyuk@uw.edu

# Landau, Ginzburg, Devonshire and others

Macroscopic modelling of ferroelectric properties refers usually to Landau-Ginzburg-Devonshire theory. This paper questions the meaningfulness of this term, discussing contributions of the three authors to what is supposed to be a theory. The limitations of every contribution are analyzed. In the main text and, to more extent in the Supplementary Material, the Landau theory is presented from an unusual perspective starting from simple mechanical models of spontaneous symmetry breaking and finishing by the Ising model. The aim of the presentation is to emphasize along with the qualitative breakthroughs the approximate character of macroscopic modelling associated with the above three authors.

Keywords: spontaneous symmetry breaking, order parameter, Landau theory

## 1. Introduction

The names of Landau and Devonshire are often mentioned in papers on ferroelectrics. Quite often one also mentions Landau-Ginzburg-Devonshire (LGD) which is more correct: it was Ginzburg [1] who, in 1945, first applied the Landau theory [2, 3] to ferroelectrics [1]. But how has this group formed? Were they doing the same though at different times? This is not a question from the science history only. There are conceptional differences between the approaches by Landau and Devonshire [4] which are useful to recall when both approaches are referred to in attempts to explain or describe experimental data. It is helpful to realize their limitations in order to not to be taken by surprise when the observed disagreements between the experiment and theory has a natural explanation: that the experimental conditions are beyond what the theory was designed for and not due to some imperfections in either the experiment or the theory.

An illustrative example is provided by the history of the findings by Minaeva et al. [5] in their study of sound propagation in uniaxial ferroelectric triglycinsulfate. The authors, inspired by theoretical results of Ref.6, were investigating dependence of sound attenuation on the angle between the polar axis and the sound propagation direction. They found that the predictions made with the use of Landau theory were qualitatively correct but in strong quantitative disagreement with the experiment. Within this paper the explanation is simple: an account for a nearly homogeneous depolarizing field excited by the acoustic wave, which was predicted to be the reason of angular dependence of the attenuation is beyond the Landau theory. This was overlooked both by the authoritative authors of Ref.6 and by their readers. This was seemingly also the case of the author of a proposal to explain this disagreement [7]. Being both qualitatively meaningful and pioneering, such an attempt was equivalent to moving beyond the Landau theory while neither acknowledging the move nor analysing such an important consequence that the theory stopped to be consistent [8]. As a result, a presumably new material constant has been introduced, the so-called 'base dielectric constant' which produced a lot of confusion including addition of this constant to the tables of material constants of ferroelectric materials. In effect, this 'constant' is nothing more than a parameter of a qualitative phenomenological model [8]. This parameter has a diffuse physical meaning and should be used with reservations and care though its order of magnitude can be, probably, realistically guessed. Anyway, it is not a material constant with a measurable numerical value as many people erroneously believed.

---

[1] We realize that usually when people mention LGD, they mean considering of gradients of polarization. This is unfortunate because Ginzburg was far from being the first to take into account this gradient term.

This unfortunate story shows that old theories, because of being old and, therefore, respected and convenient, are often used almost automatically without understanding their limitations. For this very reason, it is useful to reconsider such theories along with their conceptualities. In the same context, proposing new ways to teach them is certainly helpful for a sound development of science. The golden anniversary of the Ferroelectrics seems to be an appropriate occasion for such an undertaking, which is mainly pedagogical but also historical and methodological.

In the next Section the Landau theory is exposed from a different angle than by Landau himself. It is argued that, conceptually, this theory is closer to classical mechanics than to thermodynamics. This section is the longest in the paper and could be even longer if not being accompanied by the Supplementary Material. It is recommended to be read before and to be consulted during the reading of the main text if some of its statements look non-evident. All the figures are in the Supplementary Material where they are numerated from 1 to 16. Reference (**Sn**) is to Section n of the Supplementary Material. In Sections 3 and 4, Ginzburg's and Devonshire's approaches are critically presented. In Section 5, we comment on the role of the Landau theory in the macroscopic modelling of properties of ferroelectrics.

## 2. Landau

The Landau theory has been shown to be wrong in 1944 when Onsager published his exact solution of two-dimensional Ising model (see, e.g. Ref.3). A legend says that Landau spent a whole week trying hard to understand the Onsager paper and to check it but the Onsager treatment was so complicated that he failed and stopped trying. During many years the contradiction between Onsager and Landau was a mystery for many scientists and motivation for others to understand its reason and to put forward a new theory equally general as the Landau one but without contradictions. Landau participated in these efforts since 1958 (see Ref.3, Sec. 147) though he published nothing. Huge collective efforts resulted about 1970 in construction of a modern theory of second order phase transitions (key words: critical indices, renormalization) which had some features common with the original Landau theory (see, e.g., [9]). They are: (i) to relate the second order phase transitions to spontaneous symmetry breaking with soft stability loss, (ii) to go beyond the standard thermodynamics by introducing what is called now the order parameter, (iii) to reveal universality of asymptotic behaviour of a system close to the second order transition: this behaviour is governed by the symmetry only, not by the physical meaning of the quantities involved. The modern theory of second order transitions is, practically, not referred to in the present-day ferroelectric studies. What people mean when mentioning Landau's name is his original, old theory. This is natural and we will comment in a proper place why it is so. Here we only mention that it is this theory which is of the main interest in this paper. The main difficulty in exposing it is that, as it became clear as a result of development of the modern theory [9], the old one is not valid at the point of second order phase transition due to some thermal or quantum fluctuations. At the same time, this point is a reference point in the logic of the Landau theory. So that, to discuss its logic, it is better to begin with second order phase transitions without any fluctuations. Such transitions are possible in classical systems at zero temperature. For ferroelectric and other non-diffusional transitions this is a convenient starting point. To be illustrative we will consider mechanical models of crystals of the type often presented in textbooks: balls for atoms/ions connected by springs with whatever else that mimics interatomic/interionic forces. Since we will consider equilibrium structures only, the only branch of classical mechanics which we will refer to is statics. We will also emphasize the same Landau's achievements which were appreciated by the modern theory people. It will be argued that they are worth to be remembered in the studies on ferroelectric crystals also.

## 2.1. Spontaneous symmetry breaking

A symmetry 'breaks' when the structure loses some of its symmetry elements (operations) and converts into a structure with a smaller set of the symmetry elements (non-symmetrical structure, to be short). This is the so-called 'group-subgroup' transitions, meaning the groups of the symmetry elements. In mechanics this may happen even in a system with one degree of freedom (**S1**, **Fig. 1**). The term 'spontaneous' means that the loss of symmetry elements occurs due to change of a symmetry conserving *control parameter*, e.g. pressure or whatever else whose application conserves the symmetry properties of the system energy landscape[2] which are the reason of any symmetrical structure (The reader might find this expression vague and we refer the reader to the example below). That a symmetrical form of the energy landscape exists does not mean the system will also remain symmetrical. The loss of symmetry elements occurs because despite the symmetrical energy, the symmetrical structure becomes unstable. This happens because a minimum in the energy landscape which defined the symmetrical configuration can become a maximum or a saddle point without violating the symmetry of the energy landscape. The value of the control parameter when this happens is called its *critical* value. In the context of classical mechanical models what changes after crossing this point is the form of potential energy of the system (**S1,2 Figs. 3,4,7**). According to statics, referring to the balls & spring models considered in the Supplementary Material, the potential energy is minimum when the balls are in their equilibrium positions, i.e., any set of displacements of the balls from these equilibrium positions lead to increase of the potential energy. But after crossing the critical point the symmetrical configuration corresponds not to minimum but to a saddle point (**S2**). This occurs without violating the symmetry of the potential energy (symmetry conserving control parameter) but the system obtains at least two equivalent directions for the most effective manner of diminishing of the potential energy and, of course, it does not miss the opportunity to go along in one of these directions. As a result, the system turns into one of two or more "less symmetrical" equilibrium configurations which are equivalent with respect to the lost symmetry element(s) of the former equilibrium configuration.

---
[2] We mean energy landscape in the space of N+1 dimensions where N is the number of degrees of freedom of the system.

## 2.2. Order parameter

The initial directions of the most effective diminishing of the potential energy or the directions of the fastest decent from the saddle point are straight lines in the N- dimensional (N is the number of the degrees of freedom) space of displacements of the balls from their positions in the symmetrical configuration of the system. For N=2 this is illustrated in **S2**. These directions present patterns of the ball displacements with respect to which the stability of the symmetrical structure is lost at the critical point (**S2, S3**). Note that all the directions appear simultaneously, i.e., they are all equivalent and transform one into another at symmetry operations of the symmetrical phase. The simplest case is when there are two patterns which differ whereby the same ball displaces in opposite directions. The amplitude of the pattern shows how far the breaking of the symmetry has gone. Reinterpreting the Landau theory, we are forced to call it order parameter. This is unfortunate and paradoxical: at spontaneous breaking of symmetry in classical systems at zero temperature, i.e., in the only dominion where the Landau theory is uncontested, there is no ordering but just displacements. It would be much better to call it in another way, e.g. as 'symmetry breaking parameter'. But we will use the usual terminology believing that the reader's confusion will still be less than with a consistent but unfamiliar terminology.

In the simplest case which we will discuss in the paper we have one-component order parameter ($\eta$). The two patterns correspond to different signs of $\eta$. All the symmetry operations of the symmetrical phase either do not change $\eta$ or transform it to $-\eta$. Mathematically, this means that the order parameter transforms according to one-dimensional irreducible representation of symmetry group of the symmetrical structure. We emphasize that an exact knowledge of the pattern is not needed for the Landau theory. Its results are independent of the physical meaning of the order parameter. Nevertheless, this meaning is well defined for any specific case. Keep in mind that it treats about the amplitude of pattern of displacements (and its symmetrical equivalents) with respect to which the symmetrical structure loses its stability at the critical point. The Landau theory needs not to know this pattern to obtain its results. It needs only its transformational properties with respect to symmetry elements of the symmetrical structure. But it is well understood that this pattern is unique for every system (model). The matter is that the loss of stability occurs always with respect to a single degree of freedom. Indeed, it treats about loss of stability with respect to any, even arbitrary small perturbations, so that this loss occurs immediately when such possibility arises for a set of symmetrically equivalent degrees of freedom without waiting for arising of instability conditions for another set at a further change of the control parameter. Importantly, by defining the order parameter through the pattern of displacements with respect to which the symmetrical phase loses its stability we must acknowledge that the order parameter has a *well-defined meaning close to the critical point only*. Indeed, the pattern of structural changes resulting from the stability loss is, naturally, close to the pattern with respect to which the stability loss occurs if the control parameter is close to its critical value and the two patterns coincide when the control parameter is at its critical value, i.e. just before disappearing. This is what is called the *soft* *stability loss*. But if we move away from the critical point further into the non-symmetrical state, the structural pattern changes, there is no reason for it to remain rigid (See S3 in Suppl. Mat.).

### *2.3. Universality*

In the above arguments we did not mention any specific model so that these arguments were *model-independent, universal*. Appearance of a saddle point instead of minimum of the potential energy with respect to any set of displacements of the classical bodies (balls in our story) from their positions in the symmetrical phase is what the spontaneous symmetry breaking consist of for the considered class of systems. In the simplest case this means existence of a degree of freedom ($\eta$) such that the potential energy as a function of $\eta$ is minimum at $\eta = 0$ for the symmetrical structure but becomes maximum when the changing control parameter ($p$) crosses the critical point ($p_c$). The symmetrical structure becomes unstable but there appears the possibility of two non-symmetrical equilibrium structures whose difference with the disappeared symmetrical one smoothly increases as the control parameter continues to change after crossing the critical point. The law governing this increase should become universal as $p \to p_c$. Landau had obtained this law: $\eta_e \propto \pm\sqrt{|p - p_c|}$, the subscript 'e' means 'equilibrium'. The law was found by minimizing the universal form of the potential energy which was derived expanding the potential energy $U(\eta, p)$ into Taylor series centered at $\eta = 0$ and keeping the minimum number of the terms. The potential energies of specific systems with soft spontaneous symmetry breaking are the closer to this form the closer is $p$ to $p_c$. This form is:

$$U(\eta, p) = U_0(p) + a(p - p_c)\eta^2 + b\eta^4, \tag{1}$$

where $U_0(p)$ is unknown function and $a$, $b$ are unknown constants. For the spontaneous breaking of symmetry to be soft the constant $b$ should be positive.

Other universal laws can be obtained if we introduce the force $h$ conjugated to $\eta$. This means that the term $-h\eta$ should be added to the right-hand side of Eq. (1). It is evident, both physically and formally, that when $h \neq 0$ then $\eta_e \neq 0$ at any value of $p$, i.e., the spontaneous symmetry breaking no more exists.

The function $\eta_e(h)$ is universal at $p = p_c$ and when $h \to 0$. The latter restriction is because the universal form of the potential energy at $p = p_c$ becomes exact for $\eta \to 0$ only (the Taylor expansion!). This law is $|\eta_e| \propto |h|^{\frac{1}{3}}$. One sees that $\frac{|\eta_e|}{|h|} \propto |h|^{-\frac{2}{3}} \to \infty$, as $h \to 0$, i.e., in the universality region

$$|\eta_e| \gg |h|, \tag{2}$$

when expressed in the same dimensions. One more universal result is:

$$\frac{d\eta_e(0)}{dh} \propto |p - p_c|^{-1} \text{ as } p \to p_c. \tag{3}$$

We see that the Landau theory is *predictive*. Remarkably, its predictions are relevant for a broad class of systems. This is because of the universality: all the patterns of the same symmetry properties exhibit the same behaviour. Consider, for example, ferroelectric transition in BaTiO3 and suppose, for simplicity, that only displacements of rigid Ti- and O-ions occur at the spontaneous symmetry breaking. The 'pattern of displacements', which lead to the change of symmetry at the spontaneous symmetry breaking, is the ratio of the displacements of the two ions, or the ratio of the polarizations due to these displacements, $P_{Ti}$ and $P_O$. When studying the asymptotic behaviour at $p \to p_c$ we can fix this ratio and concentrate on the behaviour of the amplitude of the pattern. Changes in the form of the pattern, i.e. changes in the ratio are irrelevant to the asymptotic, they provide next order corrections. As we know from what is discussed above, the amplitude of the pattern, i.e., the order parameter is proportional to $\sqrt{|p - p_c|}$. Since the pattern can be considered as rigid, both $P_{Ti}$ and $P_O$ are proportional to $\sqrt{|p - p_c|}$. The same is valid for the full polarization $P = P_{Ti} + P_O$ and any other combination. That is why, referring to the universality region, one can identify the order parameter with the full polarization ignoring the real pattern with respect to which the symmetrical phase loses its stability at the spontaneous symmetry breaking. Once again, it is possible in the universality region only, not beyond it.

We have seen that the Landau theory provides several important universal results. Another matter is that, as we understand now, specific form of the universal dependencies ($\sqrt{|p - p_c|}$, $|p - p_c|^{-1}$, $|h|^{\frac{1}{3}}$) are valid for spontaneous symmetry breaking in classical systems at zero temperature only. But the revolutionary ideas of spontaneous symmetry breaking, order parameter and universality form an important part of modern theory of second order phase transitions.

The difficulties that the Landau theory experiences for non-zero temperatures, which were later overcome by the modern theory, are shortly mentioned in the Supplementary Material. We have limited ourselves by outlining how they appear when one moves from spontaneous symmetry breaking in classical mechanical systems with a few degrees of freedom and without thermal motion (zero temperature) to more realistic systems with arbitrary large number of degrees of freedom exhibiting thermal motion (non-zero temperature). First, the same models which were discussed for zero temperature are considered for non-zero temperature. For systems with finite number degrees of freedom the idea of equilibrium loses its strict meaning and what has been calledthe spontaneous symmetry breaking at zero temperature converts into another phenomenon: splitting of peak in probability density of the order parameter. Instead of the potential energy for zero temperature this phenomenon can be described by what can be called the Landau effective potential energy whose behaviour is like that of the potential energy at zero temperature but with temperature entering the model as an additional control parameter. Transition to infinite systems converts the probability densities into a single or a pair of delta-functions, which, in the case of one degree of freedom, would correspond to recovering the meaningfulness of the term 'equilibrium state' (but occurs at the limit of zero temperature only). For systems with infinite number of degrees of freedom this recovery is possible also for non-zero temperature not for any infinite system but for those of 2D or 3D. The Landau effective potential energy acquires the meaning of the Landau free energy but, according to the modern theory, it proves to be a singular function of all its variables at the point of spontaneous symmetry breaking, making the Taylor expansion impossible and rendering the Landau theory inapplicable. It was already mentioned that finding of the new asymptotic behaviour was a feat with participation of many scientists and a Nobel Prize for one of them [9]. But this new asymptotic behaviour proved to be almost of no interest for the studies of ferroelectricity. This is a separate question which we will discuss in Sec. 5.

Originally Landau did not start with zero-temperature classical system or non-diffusional transitions. Just the opposite: he begins with describing a crystalline structure in a way which considers the thermal motion. His main example of the control parameter was temperature. His first example was the atomic ordering in binary alloys, i.e. a diffusional transition. Also, he mentioned an ordering of $NH_4$ radicals in $NH_4Cl$ where a huge λ-type anomaly of specific heat has been observed and then explained by Landau two years before compiling his famous paper [2]. Unfortunately, he paid no attention to the α-β transition in quartz which was interpreted as displacive (in modern terminology) about 10 years before [10]. He could not preview, of course, that his theory would have problems just with ordering which occurs due to change in temperature and will be ideally applicable to displacive transitions in classical models at zero temperature, which he did not even mention. The only term which can be found in his paper for what we call (unfortunately) the order parameter is 'degree of ordering'. Not an impressive difference.

It is probable that the first displacive phase transition which Landau became aware of was ferroelectric transition in $BaTiO_3$ (see the next Section). It was the $BaTiO_3$ structure which was used as an illustration of a second order transition in Sec.142 of Ref.3. However, this did not substantially influence the definition of the order parameter. Here is a quotation from the second and last Landau intravital edition of the Statistical Physics.

To give a mathematical description of a phase transition of the second kind, we define a quantity $\eta$ which represents the extent to which the configuration of atoms in the less symmetrical phase differs from that in the more symmetrical phase, in the latter phase $\eta = 0$ and in the less symmetrical phase $\eta$ has positive and negative values. For example, in a transition where there is a movement of atoms (as in $BaTiO_3$) $\eta$ may be taken as the amount of displacement. [11]

Well, the displacement of what? Of Ti or of O? Or, instead, maybe some deformation of the electronic clouds? In view of the above discussion the answer is: does not matter asymptotically or while we are in the universality region where the deviations from the asymptotical behaviour are acceptable within our precision. Unfortunately, such a clarification is absent in this, otherwise magnificent, book. It sounds paradoxical but to our knowledge there is no textbook with a consistent definition of the order parameter. This inevitably led to another difficulty. Instead of the potential energy used in the above zero-T version Landau introduced what was later called 'constrained' [12] or 'incomplete' [13, 14] thermodynamic potential. It treats about thermodynamic potential of a conditionally equilibrium system. According to Landau's and Lifshitz's Statistical Physics ([3], Sec.143) they mean thermodynamic potential 'at fixed deviations from the symmetrical state, i.e. at fixed values of the order parameter'. This 'i.e.' is dubious. Given the universality, i.e. the irrelevance of the physical meaning of the order parameter to the asymptotic behaviours which results in infinite number of variants of physical meaning of the order parameter it remains unclear what to fix. It becomes meaningful if the order parameter is defined as the pattern of displacements with respect to which the more symmetrical structure loses its stability. This pattern of displacements is unique for every specific system and there is no doubt about what to fix though, within the universality region, one need to know nothing more than its symmetry properties. Beyond this region (whose boundary, recall, depends on our accepted precision) the structural changes brought to life by the stability loss cannot be characterized by a single variable except qualitatively within a model which does not claim to be exact, unlike Landau theory for zero-T classical models, which claim to be asymptotically exact.

## 3. Ginzburg

Ginzburg wrote in his memories [15] that he was impressed by the Landau theory since the pre-war times, in the very beginning of his scientific carrier. He dreamed to apply it to superfluidity and superconductivity. The obstacle was to guess what the order parameter was. It was several years after the end of the war that his idea about some wave function as the order parameter was assumed and justified by Landau. It resulted in Ginzburg's Nobel Prize more than 50 years after the idea. It is hard to imagine a person who could estimate the importance of correct choice of the order parameter better than Ginzburg. But we go now to the time well before the idea and even before the end of the war when Ginzburg became aware of ferroelectric properties of BaTiO$_3$ discovered in a neighbouring laboratory. Expectably, he applied the Landau theory to understand what ferroelectrics are and to explain the observations. He believed that the phase transition in BaTiO$_3$ was of the second order. The data were for polycrystalline samples and did not allow to establish the real order of the transition.

In 1970's, when commenting at his seminar on a talk about improper ferroelectrics where the order parameter is not polarization even in the symmetry aspect, Ginzburg remembered how in 1944 he discussed with Landau what the order parameter was for BaTiO$_3$. It was something like: "Displacements of some ions? And then we agreed that the polarization was also valid". We do not know why they agreed and if they considered the universality. We have already argued that within a simple model, where the total polarization of BaTiO$_3$ has two additive contributions, $P_{Ti}$ and $P_O$, due to the shifts of the corresponding ions, Ginzburg's choice of the order parameter as $P = P_{Ti} + P_O$ is valid. He could also say that, since the order parameter had the same transformational properties as polarization, why not to use the polarization itself. Moreover it should be introduced anyways to calculate observable quantities. So, Ginzburg's choice was natural but valid for the vicinity of phase transition only. He did not make a reservation which was natural: neither Landau made it.

Ginzburg justified the choice of polarization for the role of the order parameter in the manner frequently used until now: it is something which is absent in the non-ferroelectric phase but is present in the ferroelectric one. Excessive freedom given by this criterium was not mentioned. But the next step was consistent with the Landau theory. He wrote the 'constrained' free energy as

$$\Phi = \Phi_0 + \alpha P^2 + \frac{\beta}{2} P^4 - EP, \tag{4}$$

where $E$ is the electric field, $\Phi_0$, $\alpha$ and $\beta$ are functions of temperature and pressure. The Landau spirit is in use of a thermodynamic potential as a function of $P$ and $E$ simultaneously. In conventional equilibrium or 'unconstrained' thermodynamics one uses thermodynamic functions of external parameters whose fixation defines the equilibrium state and they are either $E$ or $P$ (another pair preferred by other authors is $E$ and $D$, we will not discuss this question that is irrelevant to this paper) Let us discuss the meaning of Eq. 4 following the previous Section, i.e. starting with zero temperature ferroelectric transition in a system governed by classical mechanics. We have a slab of the ferroelectric material with ideal electrodes connected to an exceptionally large capacitance which fixes the voltage, i.e. the electric field in the slab. Then $\Phi$ is the potential energy of this two-part system. The first three terms constitute the potential energy when the electrodes are short-circuited, so that there is no electric field and we have a transition in a mechanical system with pressure as the control parameter, i.e. what we considered in the previous Section. The entirety of Eq. 4 is the potential energy of the system when the electrodes are connected to the capacitor (for more detail see Ref.16). Its minimum provides the equilibrium, i.e., the observable, value of $P$ ($P_e$):

$$2\alpha P_e + 2\beta P_e^3 = E \qquad (5)$$

Thus, Ginzburg derived the electrical equation of state with the use of which he explained the main properties of ferroelectrics. Of course, he did not consider zero temperature transitions. Like Landau did, he considered temperature as one more control parameter which is rightful to the same extent as pressure. Anyway, Ginzburg works was a breakthrough. Using a simple mathematics, he explained the main properties of ferroelectrics. It was the beginning of what an immense number of papers was later devoted to. Now, 75 years after his first work it might be easy, as well as useful and instructive, to criticize him (together with Landau) but he was a pioneer and it is unfortunate that his name is sometimes excluded from the 'troika' by those who naively believe that his contribution is only the account for gradients of the order parameter (in the latter, by the way, he was far from being the first).

## 4. Devonshire

Devonshire studied $BaTiO_3$ to far more detail than Ginzburg. He had more experimental data to explain: monocrystals of $BaTiO_3$ were already available in the West. In particular, he considered not one- but three-component polarization vector. But we are interested in his differences with Ginzburg of other nature and will imagine that Devonshire considered one-component polarization and second order transition as well. He mentioned neither Landau theory nor Ginzburg's paper. It is unclear if he was aware of them. In principle, he could. The first Ginzburg's paper was published in English before development of the cold war with all its disastrous consequences for the exchange of scientific information between the Soviet Union and the West. But Devonshire's omission is not surprising anyway given that the Landau theory was basically neither accepted nor understood in the West until a decade later when the famous Ginzburg-Landau paper on superconductivity became known. It surprised the western scientific community by the abundance of results 'obtained from nothing', i.e., from the Landau theory applied to superconductors.

Devonshire followed the line which was begun by Mueller and Cady for Rochelle salt (see a good review by Känzig [17]). They remained within the framework of usual thermodynamics and used conventional, not 'constrained', thermodynamic functions among whose variables were either $P$ or $E$, not the both at the same time. The Mueller-Cady theory was based on an important assumption that the same expressions for the thermodynamic potentials are valid for both paraelectric and ferroelectric phases. This was a deviation from the Gibbs theory of phase equilibrium. In his theory the two phases, A and B, coexist in equilibrium if their thermodynamic potentials, $G_A(p,T)$ and $G_B(p,T)$, are equal and, importantly, $G_A(p,T)$ and $G_B(p,T)$ are considered as independent functions. The possibility to describe the properties of the both phases by a single $G(p,T,P)$ or by a more complicated, e.g., with components of the strain tensor $\sigma_{ik}$ instead of $p$, but still a single function is not evident though may seem natural. The argument [17] is that unlike Gibbs who referred to so different phases as liquid water and ice, we deal with the case where one phase is, in effect, a distortion of the other. The electric equation of state was obtained from a thermodynamic relation

$$E = \frac{\partial G}{\partial P} \qquad (6)$$

where $G(T,P)$ is a thermodynamic potential whose Taylor expansion can be written down using symmetry arguments. In the simplest case $G(T,P)$ is given by the first three terms of Eq. 4. Then using the thermodynamic formula, one obtains Eq. 5. The subscript 'e' is not needed now because only equilibrium quantities are considered to be physical so that the thermodynamic potential for the values of $P$ other than the equilibrium ones, i.e. obtained from Eq. 6 is virtually considered as a mathematical tool whose physical meaning is not specified in contrast to the Landau theory and the line followed in the Supplementary Material. The temperature dependences of coefficients $\alpha$ and $\beta$ which were taken from the Landau theory by Ginzburg, in the Mueller-Cady-Devonshire case had to be taken from the experimental data for the two phases. The form of $G(T,P)$ as a function of $P$ at different temperatures was similar to that of Figs 3, 4 or to the form of Ginzburg's free energy for $E = 0$. But unlike the latter it was for any value of $E$. When referred to the paraelectric phase (Fig. 3) all the points of the curve correspond to equilibrium states for different values of $E$. But when it treats about the ferroelectric phase (Fig. 4) Eq. 6 provides equilibrium (stable), metastable and unstable solutions for $P$ corresponding to a hysteresis loop for homogeneous polarization unrealizable in experiment but conceptionally important.

The Mueller-Cady-Devonshire approach has left many questions unanswered. One can agree that a ferroelectric phase is a distortion of the paraelectric one. But why one uses the thermodynamic variables only to describe the distortion? The changes in structure can be tricky. Does the value of ratio of displacements of O and Ti ions in BaTiO3 influence the energy? Evidently, yes. Then why only the thermodynamic, total polarization was enough for Devonshire? Yes, he was successful to fit experimental data available to him for all three ferroelectric phases of BaTiO3. But what about new data? Given that he used data for small electric field, $\varepsilon_0 E \ll P$, can we expect that his thermodynamic potential would be valid for studies such as, e.g., electrocaloric effect in very thin films or effects of homogeneous depolarizing fields mentioned in the Introduction? Since in both mentioned cases the electric field is strong, $\varepsilon_0 E \sim P$, the expectations lack a foundation. Attempts to answer, within microscopic theory, the question why Devonshire was successful [19, 20] has revealed that the success was actually not guaranteed.

## 5. Is it the Landau Theory which is popular in Ferroelectric Community?

Recall once more that the Landau theory does not claim more than an asymptotic exactness. If we add that for many-body systems it can rightfully claim this for spontaneous symmetry breaking with soft stability loss in zero-temperature classical models only one may wonder why it is so popular in the ferroelectric community. Indeed, all the ferroelectric phase transitions in perovskites are due to change in temperature and they are of the first order, i.e. they are with a hard stability loss. Nevertheless, people fit experimental data within what they call LGD theory without any care for the universality region and become surprised when the fitting is not good. Often, there will be good agreement with experiment, what is not trivial and is worth discussing.

We will discuss this question after but now consider zero-temperature first order phase transition in a classical system. Let the system be driven to the stability loss from the symmetrical phase. What happens next is a jump wise change in the system structure unlike that we considered in Sec.2 where we put the coefficient $b$ in Eq. (1) positive. Now we should put it negative but Eq. (1) becomes physically meaningless with no finite minima, i.e. with no finite equilibrium value of $\eta$. To go back to physics, we must add the next term in the $\eta$ expansion, that of the sixth order to have:

$$U(\eta, p) = U_0(p) + a(p - p_c)\eta^2 + b\eta^4 + c\eta^6 \qquad (7)$$

with $c > 0$. Have we generalized the Landau theory to first order transitions (something he never did)? No, this is not the Landau theory anymore. As it was emphasized earlier, the Landau theory is asymptotically exact when $p$ tends to $p_c$. It was so because $\eta_e(p)$ went to zero at $p \to p_c$ so that the first terms of the Taylor series become sufficient to represent the whole function (for the $\eta$ values of interest, i.e., for $\eta$ close to $\eta_e$) more and more exactly. But this is impossible for the first order transitions because $\eta_e(p_c) \neq 0$. What Eq. (6) represents is not the Landau theory. This is something superficially similar but lacking the main achievement of the Landau theory, i.e., its exactness although at the asymptotic limit and in special conditions only. Eq. (6) is useful but not as Landau theory but as a phenomenological model which provides important qualitative and reasonable quantitative results with an appropriate choice of the coefficients but it cannot claim asymptotic exactness anywhere, unlike the Landau theory. So, what is used in the ferroelectric community should be properly called as, say, Landau-like phenomenological models meaning that they use idea of order parameter (not strictly defined unlike the Landau theory), respect the symmetry in the same way as Landau did, consider temperature as a control parameter and pay no attention to inapplicability of the Landau theory to a soft spontaneous symmetry breaking in many-body systems at finite temperatures (**S6**).

But why are these models so successful? Why did Devonshire managed to describe so well temperature-driven transitions in BaTiO3 in a wide temperature range? This question is asked since long ago [18, 19] and an exhaustive answer is lacking. Some general ideas can be formulated, however. We should ask ourselves what 'a wide temperature range' means? Wide comparing with what? An answer clearly formulated in Ref.10 but going down to several previous authors is that this 'what' is the so-called, 'atomic temperature', i.e., $10^4 - 10^5$ K. This is a natural scale of temperature dependence of crystal where there are no semi-independent, i.e. weekly coupled with the rest, ions or radicals which are ordered at zero temperature and become disordered at relatively low temperatures about the energy of this weak coupling. So how the phase transitions occur at much lower temperatures as it happens in perovskites? The answer is: by chance or due to mutual cancelling of two large (atomic!) contributions specific to the perovskite structure. This idea was virtually put forward by Slater in 1950 [20] and was supported repeatedly including recent observations [21] of BaTiO3 films on substrate providing misfit strain about 1% and leading to an increase in the temperature of the ferroelectric phase transition by about 500 K. This is naturally explainable by the strain-modified cancellation of the above mentioned 'atomic' contributions. So that the temperatures of order of $10^2 - 10^3$ K are relatively low and not far from 0 K where the Landau theory is uncontested though not for a hard but for a soft spontaneous symmetry breaking. However, if the 'hard' is not too hard but is not extremely far from being soft then reality might be also not too far from what Landau predicted. This argument has been voiced by Ginzburg in 1949 [22] and is valid now also. Low (in the above meaning) temperatures of the ferroelectric transitions may be also the reason why the inapplicability of the Landau theory close to the critical point does not reveal itself [19].

From another side, Devonshire's success should not be overestimated. It was impressive, of course, that all three ferroelectric phases came out from temperature dependence of a single coefficient. But trying to achieve not only qualitative but also quantitative agreement, he was forced to suppose a temperature dependence of another coefficient with the scale of this temperature dependence being much smaller than the atomic temperature. This has incited suspicion of Vaks [19] who tried to explain this temperature dependence by the failure of Landau theory close to the point of spontaneous symmetry breaking at non-zero temperature. However, in Ref. 23 it was proposed to conserve terms up to eighth order in polarization instead of sixth order as in the original Devonshire expansion. It was found that the temperature dependence of the coefficient which Vaks was worried about was unnecessary to assume. Then additional experimental data were analysed in Ref. 24 to conclude that both eighth power terms and the temperature dependence of the famous coefficient should be considered. This process seems unlimited. The Devonshire thermodynamic potential will change unlike the Landau theory which can be rethought and taught in different ways but will explain and predict the same basic things which it did in 1937.

## 6. Conclusion

One of the widespread ways to answer the question what the Landau theory consists in is saying that it is the mean field approximation for the Ising model. This paper is a protest against this saying. It is correct that the mean field approximation for the Ising model results in the Landau theory but it is not correct that this theory consists in this approximation. The Ising model whose role in the development of theory of phase transitions cannot be overestimated is a model of order-disorder phase transitions, i.e. the transitions which are driven by changes in temperature only. We argued that the most appropriate way to present the Landau theory is not from the order-disorder but from the displacive perspective. Unlike the order-disorder transitions the displacive ones are possible at zero temperature when they are driven by change in pressure of another non-thermal control parameter. When they are of the second order and are considered for classical models the Landau theory is correct without any reservation. Thus, discussion of such transitions is the most natural starting point for teaching the Landau theory. They are also natural starting point to understand the essence of the Landau-like modelling of the ferroelectric properties. Indeed, nowadays almost all the ferroelectric studies are concentrated on perovskites or similar materials. These transitions are mainly displacive (no transition is purely displacive if occurs at non-zero temperature) and their temperatures are low comparing with the natural scale of temperature dependences in this type of materials which is the so-called 'atomic temperature' ($10^4$ - $10^5$ K). Thus, the displacive perspective is much closer to the reality for the perovskite modelling than the order-disorder one.

**Acknowledgements** AL is grateful to J.S. Young for his permission to use parts of a common unfinished work in the Supplementary Material, to B.A. Strukov for numerous useful discussions and to S.S. Krotov for his references to papers on mechanical models of spontaneous symmetry breaking. He is also grateful to Sabanci University (Istanbul, Turkey) for allowing him to teach a postgraduate course in the Fall Semester of 2010 when some of pedagogical ideas presented in this paper were born, discussed, and elaborated.

# Supplementary Material

for

Landau, Ginzburg, Devonshire and others

## Spontaneous symmetry breaking illustrated by mechanical models

### 1. One degree of freedom, zero temperature

Illustrating spontaneous symmetry breaking with the help of mechanical models is a tradition [1-5]. We present here another simple mechanical model which is more convenient for our aims than what we know from literature. Imagine a horizontal rod penetrating a ball that can move along the rod without any friction between the two (**Fig. 1**). This lateral movement of the ball is the only way the system can change its state. We shall denote the single coordinate of the ball, $v$. There is an upright spring attached to the ball and we can change the position of the upper end of the spring by moving it up ($u$ increases) or down ($u$ decreases). It is only ever moved vertically then $u$ is a symmetry conserving control parameter because its value is completely under our control, while $v$ is an internal parameter of the system that sets itself. With the spring upright and $u = 0$, the spring is relaxed, i.e. it is neither stretched, nor compressed.

Now we shall look at how the equilibrium configuration of the system depends on the value of $u$. If we set $u > 0$, the ball will be in equilibrium at $v = 0$, as indicated in **Fig. 1**. If we set $u < 0$, attempting to compress the spring, the ball will not remain at $v = 0$ and will instead come to another position of equilibrium, where the spring is relaxed, at a certain horizontal distance from $v = 0$. This position could equally be to the left or to the right of $v = 0$ as shown in **Fig. 1(b)**.

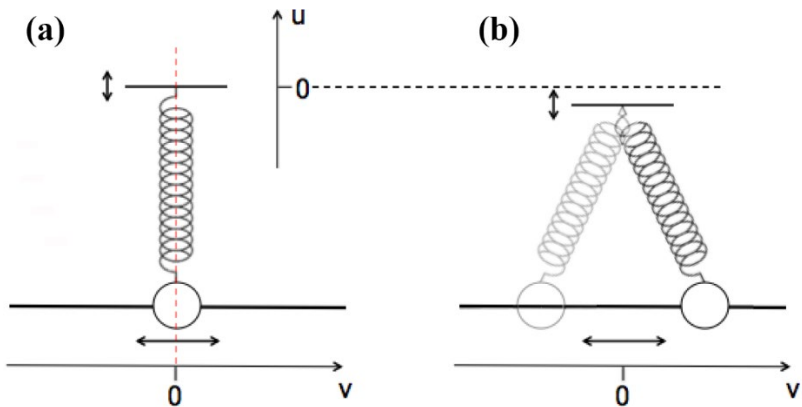

**Fig. 1** One ball and one spring. **(a)** at $u = 0$ and **(b)** for $u < 0$. Dashed red line in **(a)** indicates the position of the mirror plane.

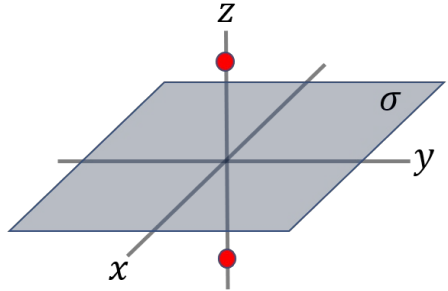

**Fig. 2** Mirror plane $\sigma$.

This is an example of 'spontaneous symmetry breaking'. For $u > 0$, the equilibrium configuration of the system does not change after a coordinate transformation known as 'mirror plane reflection'. This transformation (or 'operation') is defined in the following way in a 3D Cartesian system with axes $x$, $y$ and $z$: if the plane is situated in the $xy$ plane then the 'mirror plane reflection' will consist of transforming the coordinates of every point such that $z$ becomes $-z$, while $x$ and $y$ coordinates remain unchanged, **Fig. 2**. In our case, the relevant plane is perpendicular to the rod and positioned at $v = 0$. The transformation is then from $v$ to $-v$. The reflection operation is frequently denoted as $\sigma$, i.e. $\sigma v = -v$. A 'symmetry operation' refers to an operation that does not change the configuration of the system. The mirror plane reflection is a symmetry operation (element) for the equilibrium configuration of our system when $u > 0$, which is not the case for $u < 0$, as the mirror plane then transforms the left equilibrium position into the right one and vice versa. The system adopts a single, symmetrical configuration for $u > 0$ and one of two symmetry broken (SB) configurations when $u < 0$. As we move from $u > 0$ to $u < 0$, the system changes *continuously* between equilibrium configurations (recall that there is no static friction between the rod and the ball). This 'continuous symmetry breaking' is 'spontaneous' because we only ever move the upper end of the spring symmetrically, yet the response of the system is to take on an incongruous, less symmetrical (symmetry broken) equilibrium configuration.

Changes in equilibrium configuration result from changes in the form of the potential energy as a function of the parameters describing the configuration, i.e. $v$ in our case. For $u > 0$, the form of $U(v)$ is simple to conceive. If the ball is at $v = 0$ the spring is stretched. It is stretched even more if $v \neq 0$ and $U(v)$ increases the further the ball moves from $v = 0$ ($U(v) - U(0) \propto v^2$ for small $v$). This is shown in **Fig. 3**.

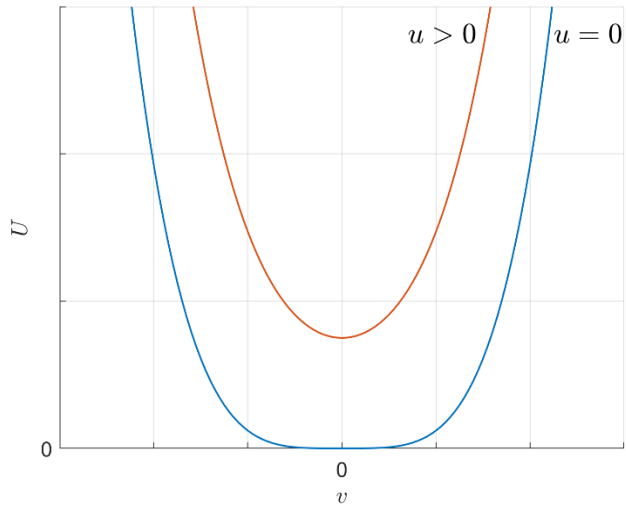

**Fig. 3** Variation of potential energy with $v$, at $u = 0$ (blue curve) and $u > 0$ (red curve).

For a given value of $u < 0$, the spring is compressed when the ball is at $v = 0$, but less compressed if we consider positions slightly to the left or right of $v = 0$. The spring finds itself more relaxed, the further the ball is situated from $v = 0$, until completely relaxed when the ball is located at one of the two possible equilibrium positions, equidistant from $v = 0$. These positions can be found considering that the spring is relaxed when it is at equilibrium. If the spring is relaxed, we say it has length $l$. In general, for a given $u$ and $v$ the length of the spring is $\sqrt{(l+u)^2 + v^2}$. From the condition $l^2 = (l+u)^2 + v_e^2$ we find that $v_e = \pm\sqrt{-2lu - u^2} \cong \sqrt{-2lu}$ for small values of $|u|$, $|u| \ll l$. At positions further from the equilibrium positions, extension of the spring qualitatively replicates the $u > 0$ scenario and **Fig. 4** describes this graphically.

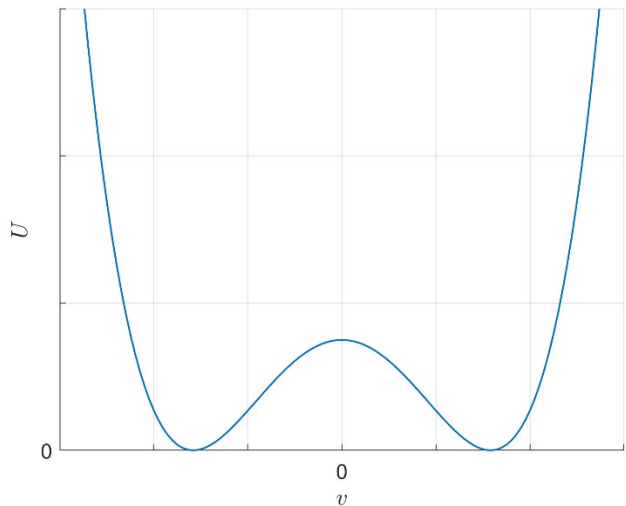

**Fig. 4** Variation of potential energy with $v$, for $u < 0$. The minima correspond to the relaxed states of the spring.

Having understood qualitatively the form of potential energy as a function of $v$ for either positive or negative values of $u$ we shall obtain now $U(v; u)$ analytically. The potential energy of the spring is

$$U(v; u) = \frac{k}{2}\left(\sqrt{((l+u)^2 + v^2)} - l\right)^2 = \frac{k}{2}\left((l+u)^2 + v^2 + l^2 - 2l\sqrt{((l+u)^2 + v^2)}\right) \quad (1)$$

This is an exact formula, but as we have already mentioned, our discussion of the mechanical models is aimed as an illustration of a more general theory. The latter claims to reveal only limiting behavior of various quantities for values of the control parameter (our $u$) close to the value corresponding to the spontaneous symmetry breaking (zero in our case) and that is why we are interested not in an exact formula for the potential energy but rather in an approximate one valid for $u$, $v$ close to zero. To obtain such a formula we first limit ourselves by small $v$ expanding the square root in terms of $v^2$ and truncating it,

$$\sqrt{(l+u)^2 + v^2} = (l+u)\sqrt{1 + \frac{v^2}{(l+u)^2}} = (l+u)\left(1 + \frac{1}{2}\frac{v^2}{(l+u)^2} - \frac{1}{8}\frac{v^4}{(l+u)^4}\right) \quad (2)$$

and therefore, obtaining

$$U(v; u) = \frac{k}{2}\left(u^2 + \frac{u}{l+u}v^2 + \frac{1}{4}\frac{lv^4}{(l+u)^3}\right) \quad (3)$$

As the next step we take into account that $l + u \simeq l$ for $u$ close to zero so that the final formula of our interest is

$$U(v; u) = \frac{k}{2}\left(u^2 + \frac{u}{l}v^2 + \frac{1}{4}\frac{v^4}{l^2}\right) = \frac{k}{2}u^2 + \frac{a}{2}v^2 + \frac{b}{4}v^4 \quad (4)$$

where

$$a = \frac{ku}{l}, b = \frac{k}{2l^2} \quad (5)$$

**Fig. 5** demonstrates that a further simplification of **Eq. (4)** (by omitting the fourth order, 'anharmonic' term) captures the conversion of minimum at $v = 0$ into maximum, i.e. the loss of stability of the symmetrical configuration but does not reveal the nonsymmetrical equilibrium configurations

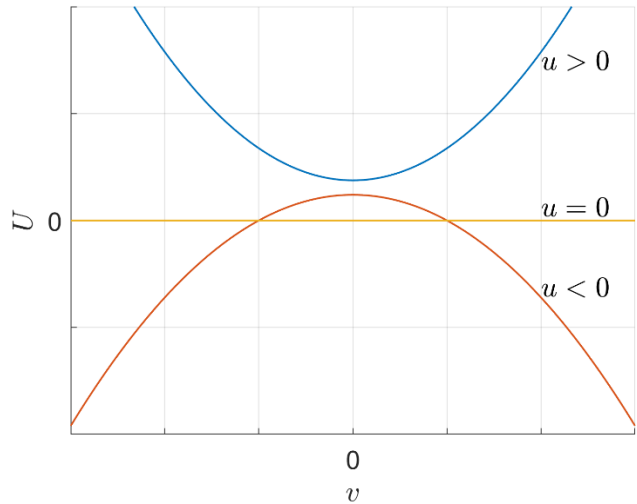

**Fig. 5** Plot of potential energy $U$ at $b = 0$. For $u < 0$ (red line) there is no minima, i.e., no equilibrium state is possible. In case of $u > 0$ (blue line) minimum is located at $v = 0$. For $u = 0$ there is a neutral equilibrium and any equilibrium value of $v$ is possible, cf. the blue curve in **Fig. 3**.

Recalling the expansion and the neglections made at obtaining **Eqs. (2-4)** one realizes that **Eq. (4)** can be considered as a good approximation if $|v|, |u| \ll l$. From the condition of minimum of the potential energy of **Eq. (4)** we obtain the same value of equilibrium values of $v$ as from the condition of the relaxed spring for $|u| \ll l$ (see the second line above **Fig. 4**).

## 2. Two degrees of freedom, zero temperature

We now consider a system with two balls on the same rod. Each one of the balls is supplied by vertical springs and there is a horizontal spring which is neutral when the vertical springs are stretched, see **Fig. 6**.

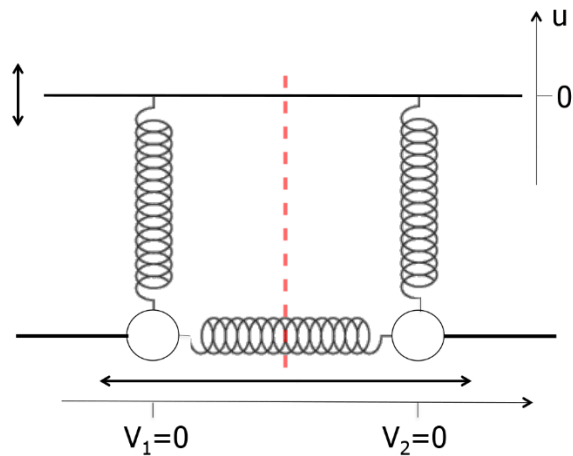

**Fig. 6** Two balls and three springs. Dashed red line indicates the position of the mirror plane.

The displacements of the upper end of the vertical springs are always identical and we shall denote it $u$ as before. Once again this is a control parameter. The displacement of the balls with respect to their equilibrium positions with stretched springs we denote as $v_1$, $v_2$. They are internal parameters and the degrees of freedom. When the springs are stretched the equilibrium configuration of the system is symmetrical with respect to the mirror plane containing the middle point between the balls. Trying to compress the springs we displace both of the balls to the left or to the right without changing the length of the horizontal spring. Thus, the dependence of the equilibrium ball displacements on $u$ is the same as in the previous Section.

Despite the system has two degrees of freedom the symmetry breaking is described by a single variable, just as in the one-ball case. This is clear from the observation that the both balls shift to the same distance to stay in equilibrium. The pattern of the displacements with respect to which the symmetry breaks is given by $v_1 = v_2$. For this system the pattern of displacements does not change with change of the control parameter but this is not always so as we will see in **Section 3**.

Let us make the above conclusion about the pattern of displacements formally considering the potential energy of the system. Once more it is a spring energy but not of a single spring but of the three ones. As to the energy of a vertical spring we have already a simplified expression which is valid in the vicinity of the symmetry breaking point ($u = 0$), see **Eq. (4)**. Since here we are also interested in this vicinity only, we shall use this expression. The change of the length of the horizontal spring is equal to $v_1 - v_2$ and therefore

$$U(v_1, v_2; u) = ku^2 + \frac{a}{2}(v_1^2 + v_2^2) + \frac{b}{4}(v_1^4 + v_2^4) + \frac{k_h}{2}(v_1 - v_2)^2, \qquad \text{(6a)}$$

or

$$U(v_1, v_2; u) = ku^2 + \frac{(a+k_h)}{2}(v_1^2 + v_2^2) + \frac{b}{4}(v_1^4 + v_2^4) - k_h v_1 v_2, \qquad \text{(6b)}$$

where $k_h$ is the spring constant for the horizontal spring.

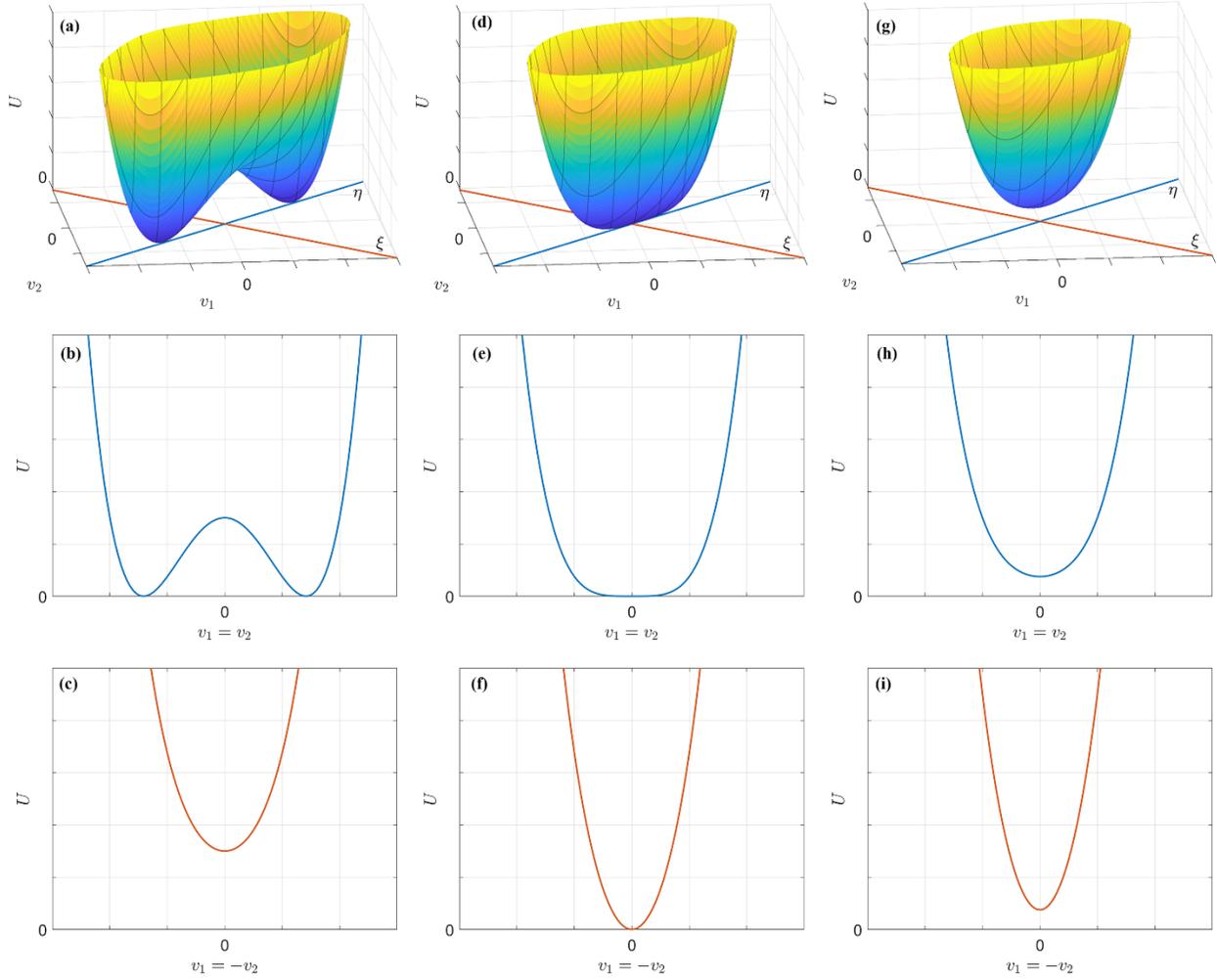

**Fig. 7 (a)**, **(d)** and **(g)** Surface plot of $U$ given by **Eq. (6b)** for $u < 0$, $u = 0$ and $u > 0$, respectively. $U$ profiles along the $\eta$ axis ($v_1 = v_2$ line) of **(a)**, **(d)** and **(g)** are shown, respectively, in **(b)**, **(e)** and **(h)**. $U$ profiles along the $\xi$ axis ($v_1 = -v_2$ line) of **(a)**, **(d)** and **(g)** are given, respectively, in **(c)**, **(f)** and **(i)**.

To obtain analytically the pattern of displacements with respect to which the loss of stability occurs and the minimum of the potential energy at $v_1 = v_2 = 0$ converts into a saddle point (**Fig. 7**) we need only a few of the terms in **Eq. (6b)**, specifically those of second order in $v_1$ and $v_2$, cf. **Fig. 5** (*harmonic part* in a frequently used jargon),

$$U_{harm}(v_1, v_2; u) = \frac{(a+k_h)}{2}(v_1^2 + v_2^2) - k_h v_1 v_2 \tag{7}$$

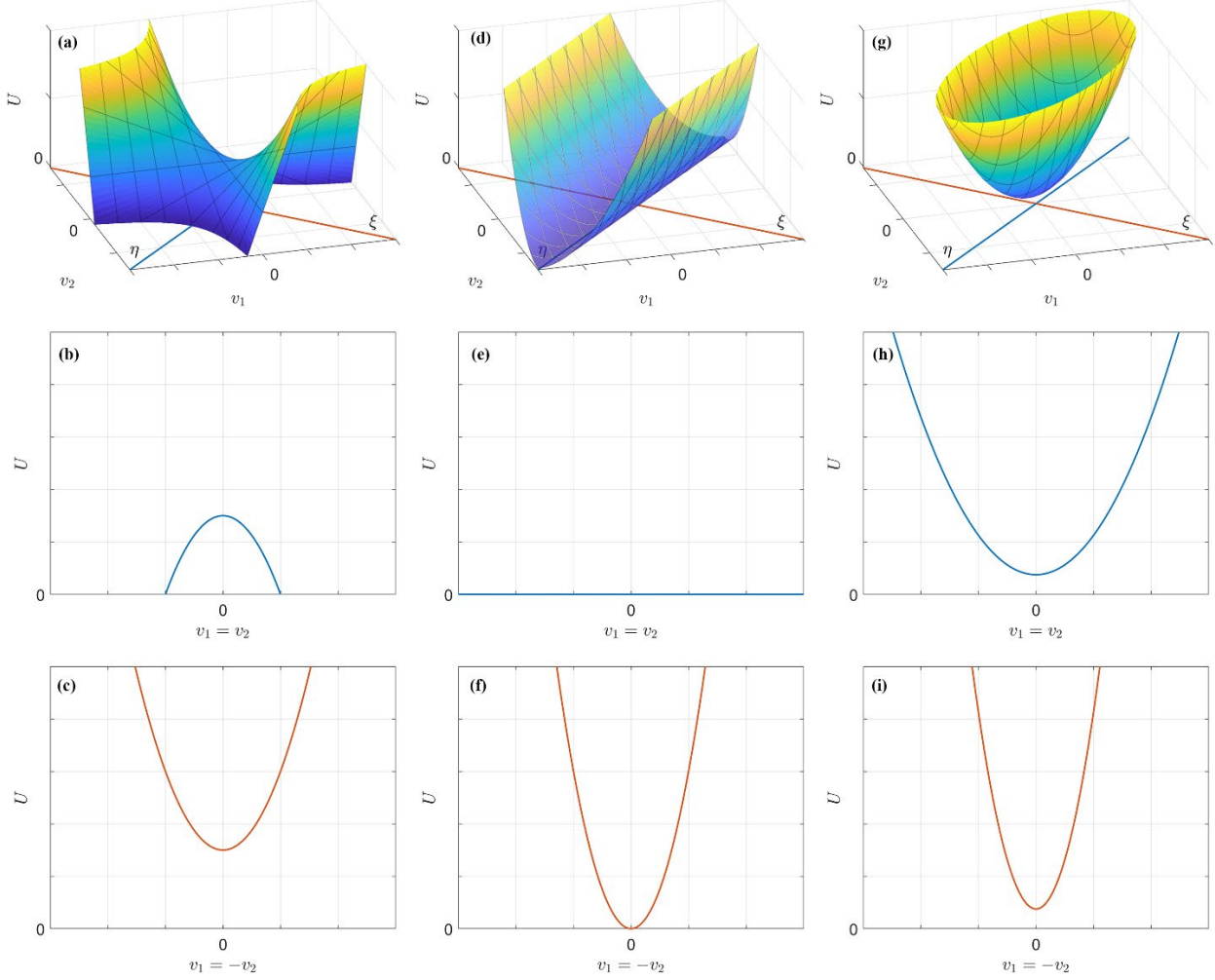

**Fig. 8 (a)**, **(d)** and **(g)** Surface plot of $U$ given by **Eq. (7)** for $u < 0$, $u = 0$ and $u > 0$, respectively. $U$ profiles along the $\eta$ axis ($v_1 = v_2$ line) of **(a)**, **(d)** and **(g)** are shown, respectively, in **(b)**, **(e)** and **(h)**. $U$ profiles along the $\xi$ axis ($v_1 = -v_2$ line) of **(a)**, **(d)** and **(g)** are given, respectively, in **(c)**, **(f)** and **(i)**.

We have seen in **Section 1** that the harmonic part of the potential energy depending on one variable converts at the stability loss into a constant as an intermediate form between the parabola up and the parabola down having either minimum or maximum at $u > 0$ and $u < 0$ (**Fig. 5**). If the harmonic part were the total potential energy, this state would correspond to neutral equilibrium. Now this constancy of the harmonic part of the potential energy ('neutral equilibrium') is expected along a line in the $v_1, v_2$ plane (**Fig. 8**). For every point on this line we have $\frac{\partial U_{harm}}{\partial v_1} = \frac{\partial U_{harm}}{\partial v_2} = 0$ or

$$(a + k_h)v_1 - k_h v_2 = 0 \tag{8a}$$

$$-k_h v_1 + (a + k_h)v_2 = 0 \tag{8b}$$

This system of two homogeneous equations has nontrivial solutions only if

$$(a + k_h)^2 - k_h^2 = a(a + 2k_h) = 0 \tag{9}$$

Of the two solutions of this equation, $a = 0$ and $a = -2k_h$, the first is relevant and the second is not. Indeed, starting from the symmetrical configuration as we decrease $a$ the stability is lost at $a = 0$ and upon a further reduction in $a$ a new, nonsymmetrical configuration forms, the symmetrical one no more exists so that when $a$ reaches the value $-2k_h$ there is no more a symmetrical configuration for which loss of stability would occur at this $a$ if this configuration existed. Now we come to an important point, at $a = 0$ **Eqs. (8a)** and **(8b)** become equivalent and both of them say the same: $v_1 = v_2$. This is the equation of the line whose all points correspond to the 'neutral equilibrium' as in **Fig. 5**. If we associate this line as a new coordinate axis the variable changing along this line is analogous to $v$ of **Section 1**. We call this variable the order parameter ($\eta$). The second axis is perpendicular to the first one and is the line $v_1 = -v_2$ (**Fig. 7 (a)**, **(d)** and **(g)**). The second variable is not important enough to have a special name (as we shall see below) and we denote it $\xi$. The new variables, which appeared due to the rotation of coordinate axes in the plane $v_1, v_2$ can be expressed through the old ones and vice versa, see **Fig. 9**.

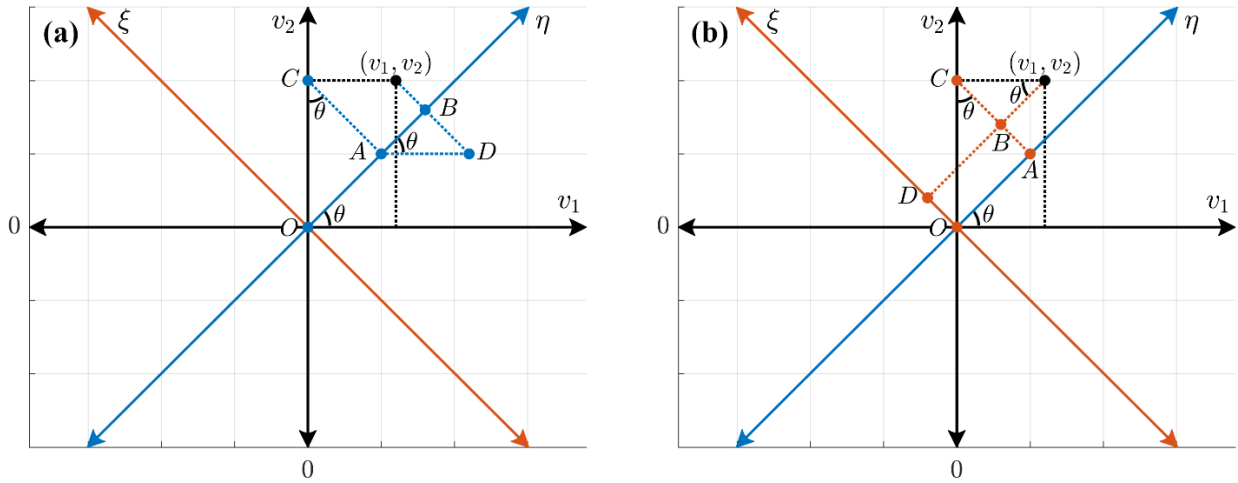

**Fig. 9** Coordinate values of a point $(v_1, v_2)$ **(a)** along the $\eta$-axis $|OB| = |OA| + |AB| = v_2 \sin\theta + v_1 \cos\theta$ and **(b)** along the $\xi$-axis $|OD| = |AB| = |AC| - |BC| = v_2 \cos\theta - v_1 \sin\theta$.

Given that $\theta = \pi/4$ one finds

$$\eta = \frac{v_1 + v_2}{\sqrt{2}}, \quad \xi = \frac{-v_1 + v_2}{\sqrt{2}} \tag{10a}$$

$$v_1 = \frac{\eta - \xi}{\sqrt{2}}, \quad v_2 = \frac{\eta + \xi}{\sqrt{2}} \tag{10b}$$

We also shall choose the new variables using symmetry arguments. From the point of view of the mirror plane reflection the main difference between the problems with one and two degrees of freedom is that the single degree of freedom transforms into itself with the opposite sign ($\sigma v = -v$) while for the variables $v_1$ and $v_2$ we have $\sigma v_1 = -v_2$ and $\sigma v_2 = -v_1$, see **Fig. 10**.

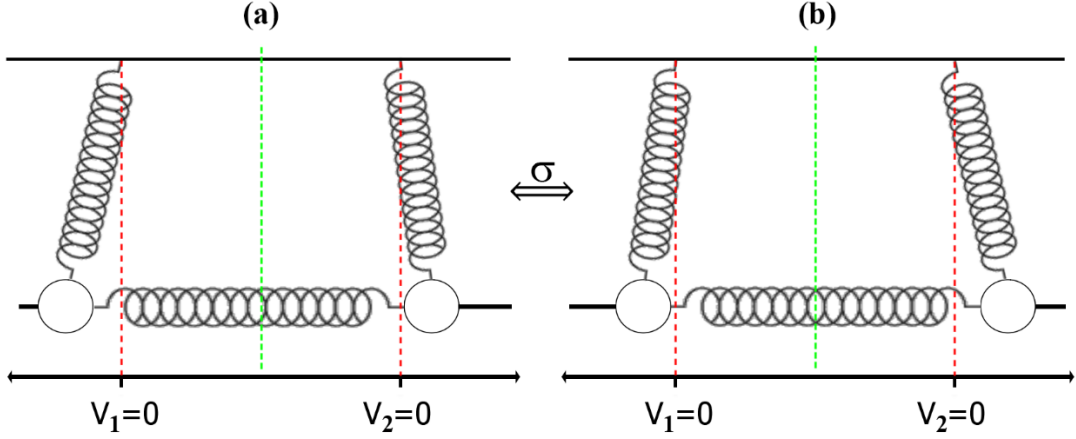

**Fig. 10.** The mirror plane transformation of a general (non-equilibrium) configuration. The transformed configuration has the same potential energy. Dashed green lines indicate the position of mirror planes.

Can we construct from $v_1$ and $v_2$ a variable with the same transformation properties as $v$? Yes, it is $v_1 + v_2$ with, of course, any factor: $\sigma(v_1 + v_2) = \sigma v_1 + \sigma v_2 = -v_2 - v_1$, i.e the variable $v_1 + v_2$ transforms into itself with the opposite sign just as $v$. Additionally, another new variable with a similar property of transforming into itself but this time without a change of sign is identified as $v_1 - v_2$. A vector in the direction of the line has equal components, i.e. the unit vector $\boldsymbol{i}_\eta = \left(\frac{1}{\sqrt{2}}, \frac{1}{\sqrt{2}}\right)$ and an arbitrary vector is $\eta \boldsymbol{i}_\eta$. In the same way for the second variable, we define a line with the unit vector $\boldsymbol{i}_\xi = \left(\frac{-1}{\sqrt{2}}, \frac{1}{\sqrt{2}}\right)$ and an arbitrary vector is $\xi \boldsymbol{i}_\xi$. The new variables $\eta$ and $\xi$ are convenient because we can be sure that the potential energy does not contain a term $\eta \xi$: $\sigma \eta \xi = -\eta \xi$, i.e., the form of the potential energy containing such a term changes after the mirror plane reflection. Therefore, the variables which we have chosen by their transformation properties at the same time *diagonalize* the second order form given in **Eq. (7)** where unlike of $\eta \xi$, the term $v_1 v_2$ is allowed by the symmetry: $\sigma(v_1 v_2) = v_1 v_2$. Any point in the plane is characterized by its radius vector which can be written down in terms of the components along the axes either $v_1$, $v_2$ or $\eta$, $\xi$, i.e.

$$v_1 \boldsymbol{i}_{v_1} + v_2 \boldsymbol{i}_{v_2} = \eta \boldsymbol{i}_\eta + \xi \boldsymbol{i}_\xi \tag{11}$$

and dot multiplying first by $\boldsymbol{i}_\eta$ and $\boldsymbol{i}_\xi$ and then by $\boldsymbol{i}_{v_1}$ and $\boldsymbol{i}_{v_2}$ we find **Eqs. (10a)** and **(10b)**. Substituting **Eq. (10b)** into **Eq. (7)** we obtain

$$U_{harm}(\eta, \xi; u) = \frac{a}{2}\eta^2 + \frac{a+2k_h}{2}\xi^2 \tag{12}$$

One sees from **Eq. (12)** that when $a$ becomes negative the minimum at the origin becomes a saddle point, i.e., maximum along the $\eta$ axis and minimum along the $\xi$. Additionally, $\xi$ is irrelevant to the problem since at $a = 0$ the minimum with respect to $\eta$ (for $a > 0$) is about to convert into a maximum (for $a < 0$) while the minimum with respect to $\xi$ remains well defined. This is, of course, expected due to the symmetry properties of the variables: $\eta$ is a symmetry breaking variable (order parameter) as $v$ in **Section 1** whereas $\xi$ is a symmetry conserving variable. Therefore, when discussing the loss of stability with respect to the spontaneous symmetry breaking, one does not have to worry about changes described by $\xi$ (there is no 'danger of escape to a lower energy' along this direction) and we can set $\xi = 0$ as it was from the beginning in the reference symmetric state whose stability we are discussing.

Substituting **Eq. (10b)** into **Eq. (6b)** we obtain:

$$U = ku^2 + \frac{a}{2}\eta^2 + \frac{b}{8}\eta^4 + \frac{a+2k_h}{2}\xi^2 + \frac{b}{8}\xi^4 + \frac{3b}{4}\eta^2\xi^2 \tag{13}$$

To formally prove that $\xi_e = 0$ in the non-symmetrical phase, we minimize $U$ with respect to $\xi$ to obtain:

$$0 = \xi_e\left(a + 2k_h + \frac{b}{2}\xi_e^2 + \frac{3b}{2}\eta^2\right) \tag{14}$$

Since the expression in the parenthesis is not zero for any $\xi_e$, $\eta$ until $a + 2k_h > 0$ and we are interested in the vicinity of the spontaneous symmetry breaking point, **Eq. (14)** has the only solution $\xi_e = 0$ and the problem is reduced to that of a single variable ($\eta$).

### 3. Three degrees of freedom, zero temperature

Now we consider three balls on the same rod (see **Fig. 11**). The central spring is supposed to be always stretched and its upper end does not move but the upper ends of the lateral springs are movable and one guesses that at some $u < 0$ a spontaneous loss of symmetry occurs. It is clear that all the balls will shift to the left or to the right but the displacement of the central ball will be less than those of the two extreme balls. The question we address first is the same: can we reduce the problem to the one degree of freedom case?

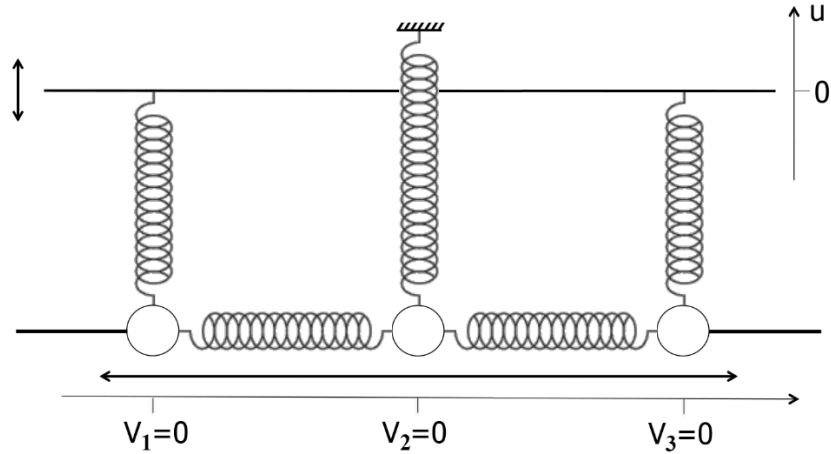

**Fig. 11** Three balls and five springs.

The potential energy of the system is

$$U(v_1, v_2, v_3; u) = \frac{3ku^2}{2} + \frac{a}{2}(v_1^2 + v_3^2) + \frac{a_1}{2}v_2^2 + \frac{b}{4}(v_1^4 + v_2^4 + v_3^4) + \frac{k(v_1-v_2)^2}{2} + \frac{k(v_2-v_3)^2}{2} \tag{15}$$

We suppose that $a_1 = k$ to simplify the formulas below without losing the conceptual aspects we want to emphasize. We also put equal the spring constants of vertical and horizontal springs. Additionally, we assume that **Eq. 15** is valid for $u$ comparable with $l$ (cf the last sentence of **Section 1**) To simplify this formula further let us look at the symmetry of the system. The mirror plane crosses the figure plane along the central line of the central spring. The symmetry operations are: $\sigma v_1 = -v_3$, $\sigma v_3 = -v_1$, $\sigma v_2 = -v_2$. Like in the previous section we see that $v_1 + v_3$ transforms as $v$ in the one-degree-of-freedom case while $v_1 - v_3$ remains unchanged. Another variable transforming as $v$ is $v_2$. We know already that the variable $v_1 - v_3$ is irrelevant to the spontaneous symmetry breaking and can be put to zero, i.e. $v_3 = v_1$ and we can consider $U$ as a function of two variables only:

$$U(v_1, v_2; u) = \frac{3ku^2}{2} + av_1^2 + \frac{k}{2}v_2^2 + \frac{b}{4}(2v_1^4 + v_2^4) + k(v_1 - v_2)^2 \tag{16}$$

We have commented above that to reveal the conditions of the spontaneous stability loss as well as its character, i.e., the pattern of the displacements realizing this loss, one can analyze the harmonic part of the potential energy.

$$U_{harm}(v_1, v_2; u) = (a + k)v_1^2 + \frac{3k}{2}v_2^2 - 2kv_1v_2 \tag{17}$$

Similar to **Section 2** we shall find the point of the stability loss and line of the 'neutral equilibrium' from the system of equations:

$$(a + k)v_1 - kv_2 = 0 \tag{18a}$$

$$-2kv_1 + 3kv_2 = 0 \tag{18b}$$

The condition for this system to have non-trivial solutions is:

$$3(a + k) - 2k = 3a - k = 0, \tag{19}$$

i.e., the stability is lost at $a = -k/3$ or at

$$u = u_c = -l/3 \tag{20}$$

It follows from either of two **Eqs. 18** at the $u_c$ that

$$v_2 = \frac{2}{3}v_1 \tag{21}$$

Recalling that $v_1 = v_3$ we obtain the pattern of the ball shifts with respect to which the stability is lost as presented in **Fig. 12**.

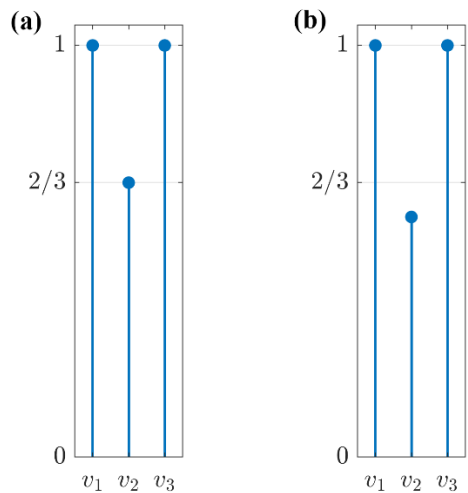

**Fig. 12 (a)** Pattern of displacements leading to stability loss. **(b)** Pattern of displacements far from the stability loss.

In the plane of $v_1$ and $v_2$ the 'neutral equilibrium' ($\eta$) line makes now an angle $\theta = \arctan\frac{2}{3}$ with the $v_1$ axis. The unit vector along the $\eta$ axis is (see **Fig. 9**) $\boldsymbol{i}_\eta = (\cos\theta, \sin\theta) = \left(\frac{3}{\sqrt{13}}, \frac{2}{\sqrt{13}}\right)$. The coordinate along the perpendicular line, $v_2 = -\frac{3}{2}v_1$, we denote as $\zeta$. This axis is with the unit vector $\boldsymbol{i}_\zeta = (-\sin\theta, \cos\theta) = \left(-\frac{2}{\sqrt{13}}, \frac{3}{\sqrt{13}}\right)$.

Presenting the same radius vector in the two forms (cf. **Eq. 11**),

$$\eta \boldsymbol{i}_\eta + \zeta \boldsymbol{i}_\zeta = v_1 \boldsymbol{i}_{v_1} + v_2 \boldsymbol{i}_{v_2}, \tag{22}$$

and dot multiplying by $\boldsymbol{i}_{v_1}$ and $\boldsymbol{i}_{v_2}$ we find:

$$v_1 = \frac{3}{\sqrt{13}}\eta - \frac{2}{\sqrt{13}}\zeta, \qquad v_2 = \frac{2}{\sqrt{13}}\eta + \frac{3}{\sqrt{13}}\zeta \tag{23a}$$

and

$$\eta = \frac{3}{\sqrt{13}}v_1 + \frac{2}{\sqrt{13}}v_2, \qquad \zeta = -\frac{2}{\sqrt{13}}v_1 + \frac{3}{\sqrt{13}}v_2 \tag{23b}$$

Substituting into **Eq. (17)** we obtain:

$$U_{harm}(\eta, \zeta, u) = a_1 \eta^2 + a_2 \zeta^2 + a_3 \eta \zeta \tag{24}$$

where

$$a_1 = \frac{9k}{13l}(u - u_c), a_2 = \frac{4k}{13l}\left(\frac{5l}{12} + u - u_c\right), a_3 = \frac{12k}{13l}(u - u_c) \tag{25}$$

The presence of the last term in **Eq. (24)** emphasizes once more what can be seen already from **Eqs (19-21)** that the change of variables according to **Eq. (23a)** diagonalizes $U_{harm}$ of **Eq. (17)** *for $u = u_c$ only*. This means that, unlike the previous section, the structural changes are described not only by the order parameter but also by another variable possessing the same symmetry properties. As a result, as one sees from **Eq. (23a)**, **Eq. (21)** is exactly valid as well for $u = u_c$ only, i.e. the pattern of displacements at the symmetry breaking point which corresponds to the order parameter coincides with the pattern of the real structural changes asymptotically close to the symmetry breaking point only. In order to elaborate on that we must substitute **Eq. (23a)** into **Eq. (16)** to obtain the full, somewhat awkward, formula for $U(\eta, \zeta; u)$. To avoid unnecessary complications we will take into account only those terms whose account implies important physical results. For that purpose, we begin with the minimum minimorum, i.e., conserving only the term proportional to $\eta^4$ to avoid absence of minima in the $\eta$-dependence and also neglect terms with $\zeta$ as we will show below that $\zeta_e \ll \eta_e$. As a result, we obtain

$$U(\eta, u) = \frac{3ku^2}{2} + a_1 \eta^2 + b_1 \eta^4 \tag{26}$$

where

$$b_1 = \frac{89}{298}b = \frac{89k}{496l^2} \tag{27}$$

Once more we come to the case of one variable finding

$$\eta_e^2 = -\frac{a_1}{2b_1} = \frac{9}{13}l(u_c - u) \tag{28}$$

Note that unlike the previous Section to come to the one-degree of freedom case, we were obliged to make an approximation which we will justify now. Obtaining this case is no more an exact procedure but asymptotically exact one. Indeed, let us also consider the terms of the lowest order in $\zeta$,

$$U(\eta, \zeta, u) = \frac{3ku^2}{2} + a_1\eta^2 + a_2\zeta^2 + a_3\eta\zeta + b_1\eta^4 + b_2\zeta\eta^3 \tag{29}$$

where

$$b_2 = \frac{84}{169}b = \frac{42k}{169l^2} \tag{30}$$

Minimizing with respect to $\zeta$ and putting $\eta = \eta_e$ according to **Eq. (28)** we find:

$$\zeta_e = -\frac{a_3\eta_e + b_2\eta_e^3}{2a_2} \cong \frac{30(u_c-u)^{\frac{3}{2}}}{l^{\frac{1}{2}}} + \cdots \tag{31}$$

We see that $\zeta_e/\eta_e \to 0$ when $u \to u_c$, i.e. can be neglected when considering the asymptotic behavior.

We also see that the variable $\eta$ does not completely describe the change of the system structure due to spontaneous symmetry breaking. It describes it the better the closer is $u$ to the symmetry breaking point ($u_c$). Recall that $\eta$ is defined as amplitude of the pattern of displacements with respect to which (the pattern) the system loses its stability at the spontaneous symmetry breaking. We have just seen that even in a system with only three degrees of freedom the pattern of the real displacements changes when the system is driven from the point of stability loss and the results of the symmetry breaking become more visible. That is why the value of amplitude of the initial pattern (the order parameter) describes the results of the spontaneous symmetry with the asymptotic precision.

**4. One degree of freedom, non-zero temperature**

Suppose now that the ball at **Fig. 1** participates in thermal movement. There are no more equilibrium positions of the ball, instead, there is an equilibrium distribution of probabilities of the ball positions given by the Boltzmann's distribution,

$$w(v, u, T) = C_1 exp\left(-\frac{U(v,u)}{k_BT}\right) = C_1 exp\left(-\frac{\frac{a}{2}v^2 + \frac{b}{4}v^4}{k_BT}\right) \tag{32}$$

where $w(v)$ is the probability density, $k_B$ is the Boltzmann's constant, $C_1$ is the normalization constant obtainable from the condition $\int_{-\infty}^{\infty} w(v)dv = 1$. The part of $U(v, u)$ which is independent of $v$ ($\frac{k}{2}u^2$ in **Eq. 4**) is included into $C_1$. The form of $w(v)$ is different at $u > 0$ and at $u < 0$, see **Fig. 13**. What now occurs at $u = 0$ is not a spontaneous symmetry breaking but what can be called *spontaneous probability peak splitting.*

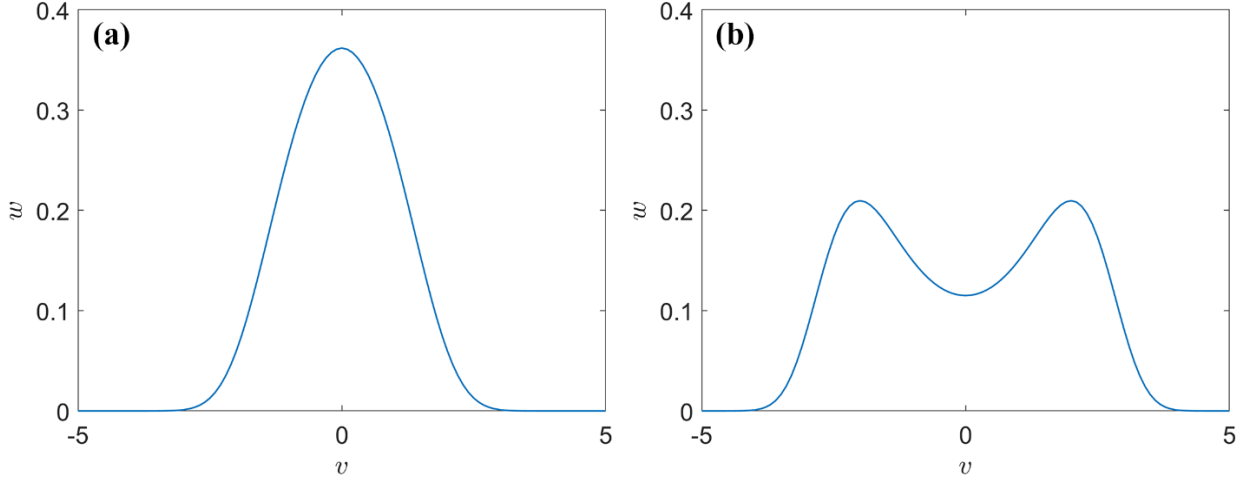

**Fig. 13** Probability density $w$ for **(a)** $u > 0$ and **(b)** $u < 0$.

As $T \to 0$ the peaks convert into delta functions and it becomes meaningful to talk about equilibrium states instead of the probability peaks.

**5. Two degrees of freedom, non-zero temperature**

Consider now the two-ball system (**Fig. 6**) at non-zero temperature. The Boltzmann's distribution can be written as:

$$w(\eta, \xi, u, T) = C_2 exp\left(-\frac{U(\eta,\xi,u)}{k_B T}\right) \tag{33}$$

To calculate $w(\eta, u)$ one should use the formula:

$$w(\eta, u) = \int_{-\infty}^{\infty} w(\eta, \xi, u) d\xi \tag{34}$$

The integral

$$\int_{-\infty}^{\infty} exp\left(-\frac{U(\eta,\xi,u)}{k_B T}\right) d\xi = exp\left(-\frac{ku^2+\frac{a}{2}\eta^2+\frac{b}{8}\eta^4}{k_B T}\right) \int_{-\infty}^{\infty} exp\left(-\frac{\frac{a+2k_h}{2}\xi^2+\frac{b}{8}\xi^4+\frac{3b}{4}\eta^2\xi^2}{k_B T}\right) d\xi \tag{35}$$

can be calculated exactly but the result is unnecessarily awkward for our aim. It makes sense to limit evaluation of **Eq. (35)** to low temperatures where ξ slightly fluctuates around zero as it attains an equilibrium value of zero at zero temperature. This justifies neglection of the $\xi^4$ term in the integrand which makes the integral familiar,

$$\int_{-\infty}^{\infty} exp\left(-\frac{\left(\frac{a+2k_h}{2}+\frac{3b}{4}\eta^2\right)\xi^2}{k_B T}\right) d\xi = \frac{\sqrt{2\pi k_B T}}{\sqrt{a+2k_h+\frac{3b}{2}\eta^2}} \tag{36}$$

Therefore,

$$w(\eta, u) = \frac{C_2\sqrt{2\pi k_B T}}{\sqrt{a+2k_h+\frac{3b}{2}\eta^2}} exp\left(-\frac{ku^2+\frac{a}{2}\eta^2+\frac{b}{8}\eta^4}{k_B T}\right) = C_2\sqrt{2\pi k_B T} exp\left(-\frac{ku^2+\frac{a}{2}\eta^2+\frac{b}{8}\eta^4+k_B T ln\sqrt{a+2k_h+\frac{3b}{2}\eta^2}}{k_B T}\right)$$

$$\tag{37}$$

We are interested in the form of $w(\eta, u)$ close to the point of spontaneous probabilities splitting, i.e., for small values of $\eta$ for which it makes sense to approximate expressions for the square root and the logarithm,

$$ln\sqrt{a + 2k_h + \frac{3b}{2}\eta^2} \cong ln\left(\sqrt{a + 2k_h}\left(1 + \frac{3b}{4(a+2k_h)}\eta^2 - \frac{9b^2}{32(a+2k_h)^2}\eta^4\right)\right)$$
$$= ln\sqrt{a + 2k_h} + \frac{3b}{4(a+2k_h)}\eta^2 - \frac{9b^2}{32(a+2k_h)^2}\eta^4 \quad (38)$$

Let us now present $w(\eta, u)$ as

$$w(\eta, u) = C_2 exp\left(-\frac{F(\eta,u,T)}{k_B T}\right), \quad (39)$$

where $F(\eta, u, T)$ we will call the Landau effective potential energy which plays the same role as the potential energy in **Eq. (32)** but depends on temperature as well.

From **Eqs. (37)** and **(38)** we find that

$$F(\eta, u, T) = F_0(u, T) + \frac{\alpha}{2}\eta^2 + \frac{\beta}{4}\eta^4 \quad (40)$$

where

$$F_0(u, T) = ku^2 + \frac{k_B T}{2}\ln\frac{a+2k_h}{2\pi k_B T} \quad (41)$$

$$\alpha(u, T) = a + \frac{3bk_B T}{2(a+2k_h)} = \frac{ku}{l} + \frac{3kk_B T}{4l(ku+2lk_h)} \quad (42)$$

$$\beta(u, T) = \frac{b}{2}\left(1 - \frac{9bk_B T}{4(a+2k_h)^2}\right) = \frac{k}{4l^2}\left(1 - \frac{9kk_B T}{8(ku+2k_h l)^2}\right) \quad (43)$$

The Landau effective potential energy converts into the usual potential energy as $T \to 0$: **Eq. (40)** converts into **Eq. (13)** with $\xi = 0$. The maxima in the probability density become minima of the potential energy and the probability density beyond the minima tends to zero. The idea of the equilibrium states is recovered.

Pay now attention to **Eq. (42)**. It shows that the spontaneous probabilities splitting occurs not at $u = 0$ as in the previous Section but at $u = u_c(T)$ obtainable from the equation $\alpha(u, T) = 0$, i.e.,

$$u_c + \frac{3k_B T}{4(ku_c+2lk_h)} = 0 \quad (44)$$

For low temperatures $u_c$ is small and can be neglected in the denominator. Therefore,

$$u_c(T) = -\frac{3k_B T}{8k_h l} \quad (45)$$

see **Fig. 14**. Note that temperature now is the second control parameter, i.e. fixing a negative value of $u$, say, at zero temperature, we can obtain, for a system of two interacting (horizontal spring) balls in two-well potentials (vertical springs), a transition from the state with split probability peak to the merged one by changing *temperature only*. It is seen from **Eq. (45)** that temperature of the splitting or merging of the peaks is:

$$T_s = \frac{8k_h l |u|}{3k_B} \qquad (46)$$

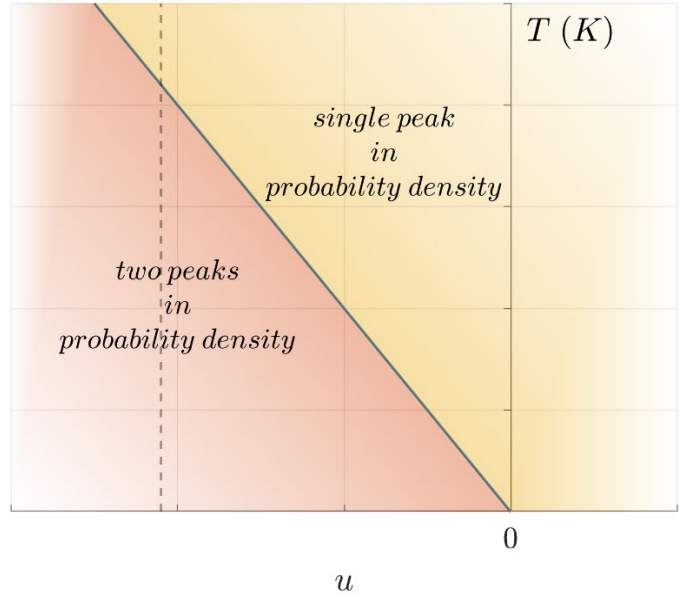

**Fig. 14** $T - u$ diagram of the spontaneous probability peak splitting. The broken line illustrates the possibility to obtain this splitting (merging) due to change of temperature only.

**6. Two balls at non-zero temperature: Lenz-Ising simplification**

When considering merging of probability peaks due to change of temperature only and starting from a fixed negative $u$ at zero temperature, one can simplify the mathematics using an idea which has been put forward in 1920 (one more jubilee) by Lenz but now is associated with the name of Ising. A body in two-well potential is idealized by Ising variable ('spin') which assumes two values, $\pm 1$, for positions of the body in the two wells. We have a system of two Ising spins $s_1, s_2$. The potential energy of this system has two values: for the same signs of $s_1, s_2$ when both balls are either in the left or in the right wells and for opposite signs of $s_1, s_2$ when one ball is in the left well and the other one is in the right well or vice versa. A convenient formula for the energy is $-I s_1 s_2$ so that when the balls are in similar wells the energy is $-I$ and when they are in dissimilar wells the energy is $I$. The difference between the energies is $2\,I$.

Let us reconsider this in terms of our two-ball model. Begin with removing the horizontal spring making the two balls independent of each other. When $u$ becomes negative each ball has two possibilities to move either to the left or to the right so that at zero temperature we have a total of four equilibrium states: both of the balls move leftward (rightward) or one ball moves leftward (rightward) while the other one moves in the opposite direction. If we have a set (an ensemble) of identical systems, then the probability of finding a state is the same for either of the four states. At non-zero temperature the balls exhibit fluctuations close to the former equilibrium positions and sometimes due to a stronger fluctuation a ball may transit from the vicinity of, say, left equilibrium position (minimum of the potential energy) to the vicinity of the right one. The Lenz-Ising simplification consist in reduction of infinite set of states available to a ball at thermal motion to two states only: those of being in the vicinity of one minimum of the potential energy or in the vicinity of the other one. These are the two possible values of the Lenz-Ising variable. In the absence of the horizontal spring the system finds itself with the same probability in any of the four possible states. One describes this situation as complete disorder. When the horizontal spring connects the two balls, as described in Sec.2, we have two equilibrium states at zero temperature but at a non-zero temperature the other two states, as well, become available to the system. Their energy is higher, i.e., the horizontal spring is either stretched or compressed as shown in (**Fig. 11**). If the spring constant of the horizontal spring is much less than of the vertical ones one can neglect the shifts of the balls from their positions without the horizontal spring. Using **Eq. (9)** we calculate that the change of the length of the horizontal spring due to either stretching or compression is $\sqrt{2}\eta$ so that the excess of energy compared with two other states is $k_h \eta^2 = 2k_h ul$ (see Sec.1), i.e., within our model $I = k_h ul$.

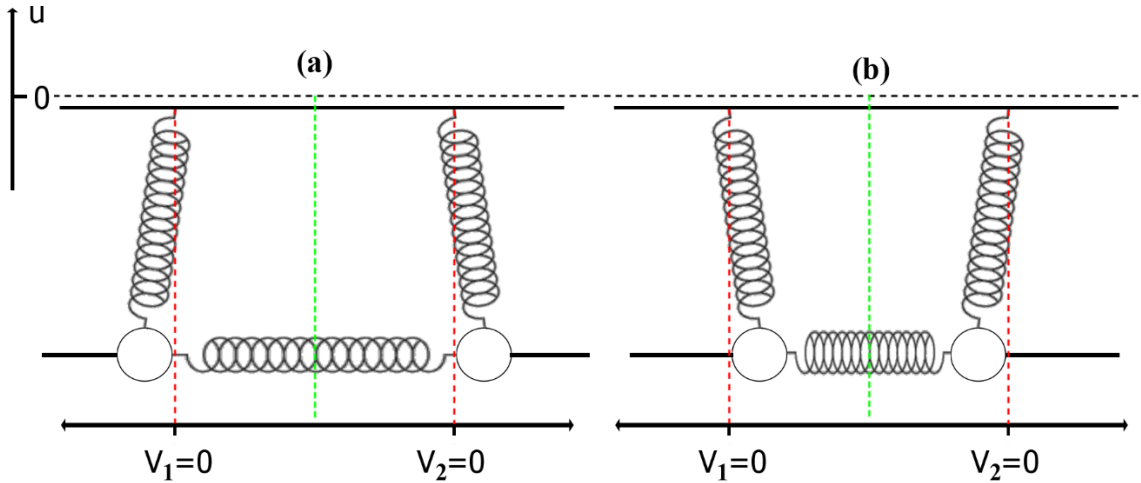

**Fig. 15** Higher energy configurations of the system when horizontal spring is **(a)** stretched, (balls move away from each other), **(b)** compressed, (balls move towards each other) for $u < 0$. Dashed green lines indicate the position of mirror planes.

Using the above idealization, it is convenient now to study the spontaneous probability splitting in systems with many equivalent balls connected by horizontal springs. Begin with a system of two balls. The potential energy is

$$U = -I s_1 s_2 \quad (47)$$

We shall argue below that now it is convenient to define the order parameter as an 'average spin', i.e,

$$\eta = \frac{s_1 + s_2}{2} \quad (48)$$

This is similar to our definition of $\eta$ above, cf **Eqs. (9)** and **(10)**. Now $\eta$ is a discrete variable assuming values -1, 0, 1. Let us calculate the probabilities of these values. According to the Boltzmann principle the probability of a state number $i$ with an energy $U_i$ is

$$p_i = C \exp(-U_i / k_B T) \quad (49)$$

where $C$ is to be found from the condition $\sum_i p_i = 1$. The probability to have one of the possible values of $\eta$ is:

$$p_{\eta_j} = \Sigma_i p_i \tag{50}$$

where the summation is over all the states with the same value of $\eta_j$. For $\eta_3 = 1$ there is only one state: (1,1), i.e., $s_1 = s_2 = 1$ but for $\eta_2 = 0$ there are 2 states: (-1,1) and (1,-1) with the same energy. We have, therefore,

$$p_{\eta_1} = p_{\eta_3} = Cexp\left(\frac{I}{k_BT}\right), \quad p_{\eta_2} = 2Cexp\left(-\frac{I}{k_BT}\right) \tag{51}$$

where

$$C = \left(2exp\left(\frac{I}{k_BT}\right) + 2exp\left(-\frac{I}{k_BT}\right)\right)^{-1} \tag{52}$$

Note that at $T = 0$ we have $p_{\eta_1} = p_{\eta_3} = 1/2$ and $p_{\eta_2} = 0$ while at $T \to \infty$ we obtain $p_{\eta_1} = p_{\eta_3} = 1/4$ and $p_{\eta_2} = 1/2$, see **Fig. 16**.

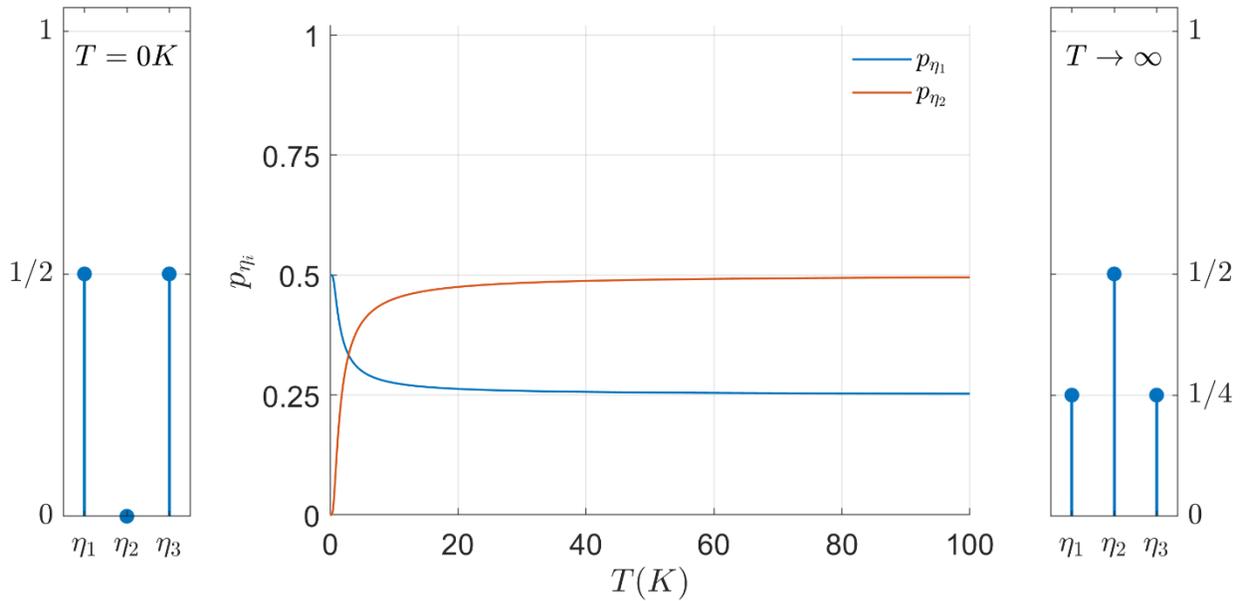

**Fig. 16** Probability as a function of temperature along with pattern of displacements at $T = 0$ K and as $T \to \infty$.

There is a clear indication on splitting of probability peak upon cooling from a high temperature as it was found in Sec.S5. The temperature of the splitting ($T_s$) is to be found from the condition $p_{\eta_1} = p_{\eta_2}$, i.e.

$$exp\left(\frac{I}{k_BT_s}\right) = 2exp\left(-\frac{I}{k_BT_s}\right) \tag{53}$$

or

$$T_s = \frac{2I}{k_B ln2} = \frac{2k_\hbar ul}{k_B ln2} \tag{54}$$

This is close to the result for the non-simplified model (**Eq. (47)**) despite of a drastic character of the simplification.

### 7. From several to infinite number of balls

Our final aim is, of course, to model real atomic system where the number of degrees of freedom is immense, recall Avogadro number. We considered systems with several degrees of freedom because some features of the phenomenon (spontaneous symmetry breaking at zero temperature and spontaneous probability peak splitting at non-zero temperatures) remain the same in systems with any number of balls but others change.

Begin with the zero temperature. Looking at **Fig. 6** it is easy to imagine an infinite rod with infinite number of balls and springs. This will be the infinite periodic model with one ball in the repeating unit (unit cell). Evidently, at $u = 0$ there will be a spontaneous symmetry breaking just as it was discussed in **Sections 1** and **2**. How to formally define the order parameter in this case so that the definition will also be valid in the case when the number of the balls, $N$, tends to infinity? Above we defined it, in effect, as the length of the vector in the $N$-dimensional space of the ball's displacements whose direction defines the pattern of displacements with respect to which the system loses its stability at the symmetry breaking. Generalizing **Eq. (11b)** we have

$$\eta = \frac{v_1 + v_2 + \cdots v_N}{\sqrt{N}} \tag{55}$$

If $v_1 = v_2 = \cdots = v_N = v$ we have $\eta = \sqrt{N}v$, i.e. $\eta \to \infty$ when $N \to \infty$. This is inconvenient. Since multiplication of a vector by a scalar does not change its direction, we can multiply **Eq. (55)** by $1/\sqrt{N}$. Designating the left-hand side once more by $\eta$ we obtain a more convenient definition

$$\eta = \frac{v_1 + v_2 + \cdots v_N}{N}. \tag{56}$$

For our model with the balls the order parameter has a clear physical meaning of average displacement of the balls. It is this definition which we virtually used in the previous Section.

Looking now at **Fig. 11** one can easily imagine another periodic system where the upper ends of the springs of the balls with even numbers are fixed while the upper ends of the springs of the balls with odd numbers are free to move. Now there are two balls in the unit cell of the model. It is clear that all the even numbered balls shift in the same way at the spontaneous symmetry breaking as well as all the odd numbered ones. To define the order parameter in terms of the ball shifts we should first define, similar to the above, the average shifts of the even- and odd-numbered balls

$$v_{even} = \frac{v_1 + v_3 + \cdots v_{2k+1}}{2k+1} \tag{57}$$

$$v_{odd} = \frac{v_2 + v_4 + \cdots v_{2k}}{2k} \tag{58}$$

and then can use **Eq. (23b)** to define the variables $\eta$ and $\zeta$ which we used in **Section 3** to describe equilibrium structure which forms as a result of spontaneous symmetry breaking. We have seen there that the symmetry breaking is associated with $\eta$, but not with $\zeta$, so that we call $\eta$ the order parameter according to its definition in the main text. Though the pattern of the displacements with respect to which the stability is lost is defined by the condition $\zeta = 0$. But this is within our model theory, not within the Landau one. The Landau theory is interested in the symmetry properties of the order parameter only and need not to know how this variable is expressed through the variables of a model theory. It also does not need the knowledge that for our model there are two variables with the same symmetry properties. Only one of them is relevant to the stability loss and it is enough to conclude that we need a single variable with definite symmetry properties to be used as the order parameter. Evidently, the potential energy increases with an increase in the number of balls. In the two periodic infinite systems, discussed above, the spontaneous breaking of symmetry is same in every unit cell so that one can consider the potential energy per unit cell.

Turning to non-zero temperatures we mention first that unlike the zero-temperature case where to describe the equilibrium structure we can consider only a few variables (one or two in the above examples) one has to consider now all the degrees of freedom of the system because all of them participate in the thermal movement and though a single variable exhibits the probability peak splitting all of them influence on where it occurs (see **Section 6**). In other words, the integral in **Eq. (34)** converts into an integral over $N-1$ variables where $N$ is the number of degrees of freedom of our model. The Landau effective potential energy is now proportional to $N$ and one has to once more consider specific Landau effective potential energy, e.g., $F(\eta; u, T) = Nf(\eta; u, T)$. Instead of **Eq. (39)** we have now

$$w(\eta; u, T) = C_\infty \exp\left(-\frac{Nf(\eta; u, T)}{k_B T}\right) \tag{59}$$

The maxima of the probability density are once more at the minima of $f(\eta; u, T)$ which are two for $u < 0$ and $T < T_c$ (at $\eta^2 = \eta_{min}^2$) and a single minimum at $\eta = 0$ otherwise (cf. **Fig. 13**). Consider a ratio between the probability densities at $\eta = \eta_{min}$ and at a neighboring point $\eta = \eta_{min} + \Delta\eta$:

$$\frac{w(\eta_{min} + \Delta\eta)}{w(\eta_{min})} = \exp\left(-\frac{N(f(\eta_{min} + \Delta\eta) - f(\eta_{min}))}{k_B T}\right) \tag{60}$$

Since $(f(\eta_{min} + \Delta\eta) - f(\eta_{min})) > 0$ the ratio tends to zero as $N \to \infty$ for an arbitrary small value of $\Delta\eta$. Recalling that the integral of $w(\eta; u, T)$ is always equal to unity we conclude that the function $f(\eta)$ has a delta function-type singularity at $\eta = \eta_{min}$. Unlike **Fig. 13** the spontaneous probability density splitting looks now as conversion of $w = \delta(\eta)$ into $w = \frac{1}{2}\delta(\eta - \eta_{min}) + \frac{1}{2}\delta(\eta + \eta_{min})$.

It may seem that now the possible values of $\eta$ are well defined and there is no way for the system to switch in the process of thermal motion between $\eta_{min}$ and $-\eta_{min}$ because the probability for $\eta$ to accept an intermediate value is zero. This would be an erroneous conclusion because what is virtually absent is the possibility of this switching via homogeneous change of $\eta$ through the infinite system. But there are possibilities of fluctuational switching via domain wall propagation which is beyond this Supplementary Material considering only states with homogeneous order parameter. This fluctuational switching makes impossible for one-dimensional Lenz-Ising model (and the most probably for our model of balls on an infinite rod) to ascribe a definite value of $\eta$ to equilibrium states for any non-zero temperature. This is what was shown by Ising in 1925. Recall that the ascription is perfectly possible for zero temperature as we have seen above.

It is also possible for two-dimensional Lenz-Ising model which was reported by Onsager in 1947. But the temperature dependence of $\eta_e$ proved to be very different from what we found in **Section 5** and what follows from the Landau theory if we carelessly (as Landau virtually did) generalize it to systems with any number of degrees of freedom. Instead of the Landau result, $\eta_e \propto (T_c - T)^{1/2}$ Onsager found $\eta_e \propto (T_c - T)^{1/8}$. This disagreement was enough to show that something is wrong with the Landau theory for atomic systems just at the reference point for the theory: the spontaneous symmetry breaking or the second order phase transition. The contradiction was resolved at the end by creating the modern theory and understand at what point the initial Landau theory was wrong. Meanwhile in the ferroelectric community the Landau theory or, rather, LGD was successfully used to describe or predict many experimental data. This provoked addressing another problem: to explain why the Landau theory or, rather, Landau-like phenomenological models are so successful. Without entering the details and meaning of the perovskites, it reduces to arguing that the ferroelectric transitions occur at low temperatures which specifically means at temperatures much lower than the so-called 'atomic temperature, $10^4 - 10^5 K$ which corresponds to characteristic energies of interatomic bonds in ionic crystals (see **Ref. 6** and the references therein). In other words, when dealing with perovskites we are not far from the zero temperature and study of the Landau-like modelling using mechanical models to understand its limitations makes both conceptual and practical sense.